\documentclass{article}

\usepackage{arxiv}

\usepackage[utf8]{inputenc}
\usepackage[T1]{fontenc}
\usepackage{natbib}

\usepackage{amsmath,amsfonts,bm}

\def\eqref#1{equation~\ref{#1}}

\def\1{\bm{1}}

\DeclareMathAlphabet{\mathsfit}{\encodingdefault}{\sfdefault}{m}{sl}
\SetMathAlphabet{\mathsfit}{bold}{\encodingdefault}{\sfdefault}{bx}{n}

\usepackage{float}
\usepackage{wrapfig}
\usepackage{capt-of}
\usepackage{hyperref}
\usepackage{url}
\usepackage{booktabs}
\usepackage{multirow}
\usepackage{array}
\usepackage{graphicx}
\usepackage{xcolor}
\usepackage{tcolorbox}
\tcbuselibrary{breakable,skins}
\usepackage{needspace}
\usepackage{xspace}
\usepackage{amsmath,amssymb}
\usepackage{pifont}
\usepackage{tikz}

\newcolumntype{P}[1]{>{\raggedright\arraybackslash}p{#1}}

\definecolor{contaccent}{HTML}{2F5D8C}
\definecolor{oursrowbg}{HTML}{E8F0F9}
\definecolor{totalrowbg}{HTML}{F2F2F2}
\definecolor{yesmark}{HTML}{2E7D32}
\definecolor{nomark}{HTML}{A6A6A6}

\newlength{\tableedgepadlen}
\newcommand{\tableedgepad}{\hspace{\tableedgepadlen}}
\makeatletter
\newcommand{\tablerowband}[1]{\leavevmode\rlap{\hspace{-\tableedgepadlen}\textcolor{#1}{\rule[-\dp\@arstrutbox]{\textwidth}{\dimexpr\ht\@arstrutbox+\dp\@arstrutbox\relax}}}}
\makeatother
\newcommand{\tableheaderrow}{}
\newcommand{\tablesubheaderrow}{}
\newcommand{\tableoursrow}{\tablerowband{oursrowbg}}
\newcommand{\tabletotalrow}{\tablerowband{totalrowbg}}

\newtcolorbox{promptbox}[1]{
  enhanced,breakable,title={#1},title after break={#1 (continued)},
  colback=white,colbacktitle=oursrowbg,colframe=contaccent!45,coltitle=contaccent!80!black,
  fonttitle=\bfseries\small,fontupper=\small,
  boxrule=0.45pt,arc=3pt,
  left=8pt,right=8pt,top=5pt,bottom=5pt,
  toptitle=4pt,bottomtitle=4pt,
  before skip=6pt,after skip=6pt
}

\newcommand{\dataset}{\textsc{Emilia2}\xspace}
\newcommand{\shortset}{\textsc{Emilia2-Short}\xspace}
\newcommand{\longset}{\textsc{Emilia2-Long}\xspace}
\newcommand{\dialogueset}{\textsc{Emilia2-Dialogue}\xspace}

\newcommand{\emiliapipe}{\textsc{Emilia2-Pipe}\xspace}
\newcommand{\yes}{\textcolor{yesmark}{\scalebox{0.85}{\ding{51}}}}
\newcommand{\no}{\textcolor{nomark}{\scalebox{0.85}{\ding{55}}}}

\title{Emilia2: A Large-Scale Dataset for Controllable Long-Form Speech Generation}

\author{\parbox{\dimexpr\textwidth-2\tabcolsep\relax}{\centering
    \mbox{Qinke Ni$^{1}$}\quad
    \mbox{Dekun Chen$^{1}$}\quad
    \mbox{Junwen Qiu$^{1}$}\quad
    \mbox{Huan Liao$^{1}$}\quad
    \mbox{Xu Li$^{1}$}\quad
    \mbox{Xingwen Han$^{1}$}\quad
    \mbox{Wan Lin$^{1}$}\quad
    \mbox{Kunyu Feng$^{1}$}\quad
    \mbox{Yuxiang Wang$^{1}$}\quad
    \mbox{Yusheng Mu$^{1}$}\quad
    \mbox{Chaoren Wang$^{1}$}\quad
    \mbox{Yingda Shen$^{1}$}\quad
    \mbox{Mingjie Chen$^{1}$}\quad
    \mbox{Haolong Zheng$^{2}$}\quad
    \mbox{Xulin Fan$^{2}$}\quad
    \mbox{Mark A.~Hasegawa-Johnson$^{2}$}\quad
    \mbox{Zhizheng Wu$^{1}$}\\[6pt]
    {\normalfont
    $^{1}$The Chinese University of Hong Kong, Shenzhen\qquad
    $^{2}$University of Illinois at Urbana-Champaign}}}

\renewcommand{\shorttitle}{Emilia2}
\date{}

\begin{document}

\maketitle

\begin{abstract}
    Generating long-form speech requires models to keep content, speaker identity, turn-taking, and expressive style consistent over time. Yet most open training corpora contain examples lasting only tens of seconds and open long-form benchmarks cover at most two speakers per recording.
    We introduce \dataset, a 229K-hour multilingual dataset
    for controllable long-form speech generation.
    From in-the-wild data, \dataset constructs short utterances, single-speaker long-form speech, and
    multi-speaker dialogues, preserving transcripts,
    speaker identities, pauses, and acoustic transitions. Examples longer than five minutes account for 50.0\% of single-speaker long-form hours and 66.8\% of dialogue hours.
    To support expressive control, \dataset provides natural language
    instructions for 4.77M examples and nonverbal-vocalization tags for 933K examples.
    We also introduce \mbox{\dataset-testset}, 23 hours of bilingual recordings of up to 30 minutes and four speakers, each speaker paired with reference audio.
    We train autoregressive (AR) and non-autoregressive (NAR) models on \dataset. Our 0.6B-backbone NAR model achieves the lowest Chinese CER and English WER on TTSD-Eval (5.27\%/9.93\%) among compared systems, while our AR model is on par with Qwen3-TTS-12Hz-VD on InstructTTSEval (zh/en: 76.2/78.3 vs.\ 77.1/77.9).
    These results validate \dataset as a comprehensive data foundation for long-form speech generation and minute-scale expressive control.
    Demo: \url{https://amphionteam.github.io/Emilia2-demo/}.
\end{abstract}

\section{Introduction}
\label{sec:intro}

Speech synthesis is moving beyond short utterances towards long-form,
human-like spoken generation \citep{nishimura2025hall,wang2026mgm,song2026borderless,zhang2026swantale}.
Applications such as audiobooks, podcasts, audio dramas, and conversational
agents require speech that lasts coherently over minutes
\citep{cheng2025dnaspeech,peng2026vibevoice,ju2026mooncast}. In these settings, controllable
long-form generation requires more than producing intelligible speech in a
target voice: it must maintain accurate content alignment and consistent
speaker identities throughout an extended recording, preserve natural pauses
and smooth transitions across utterances and speaker turns, and support
expressive control \citep{zhang2026moss,li2026swanvoice,pan2026comprehensive}.

Zero-shot text-to-speech (TTS) systems can clone a target voice from seconds of reference audio \citep{chen2025neural,anastassiou2024seed}, but most are designed for utterances of tens of seconds. Simply concatenating these short segments into longer recordings often leads to timbre drift, unnatural pauses, and prosodic discontinuities, degrading the naturalness of long-form speech. To address these challenges, recent studies explore long-form speech generation, with systems such as HALL-E and MOSS-TTSD \citep{nishimura2025hall,zhang2026moss}. Nevertheless, zero-shot long-form speech synthesis still offers limited expressive control: it remains difficult to preserve a target speaker's identity while explicitly controlling speaking style and nonverbal vocalizations (NV) such as laughter and sighs, which are essential for vivid and engaging speech. Despite rapid progress in long-form generation, large-scale long-form training corpora providing expressive annotations remain scarce. Existing corpora are often unavailable, and open benchmarks with per-speaker reference audio remain limited in duration or speaker coverage.

\begin{figure}[t!]
    \centering
    \includegraphics[width=\textwidth]{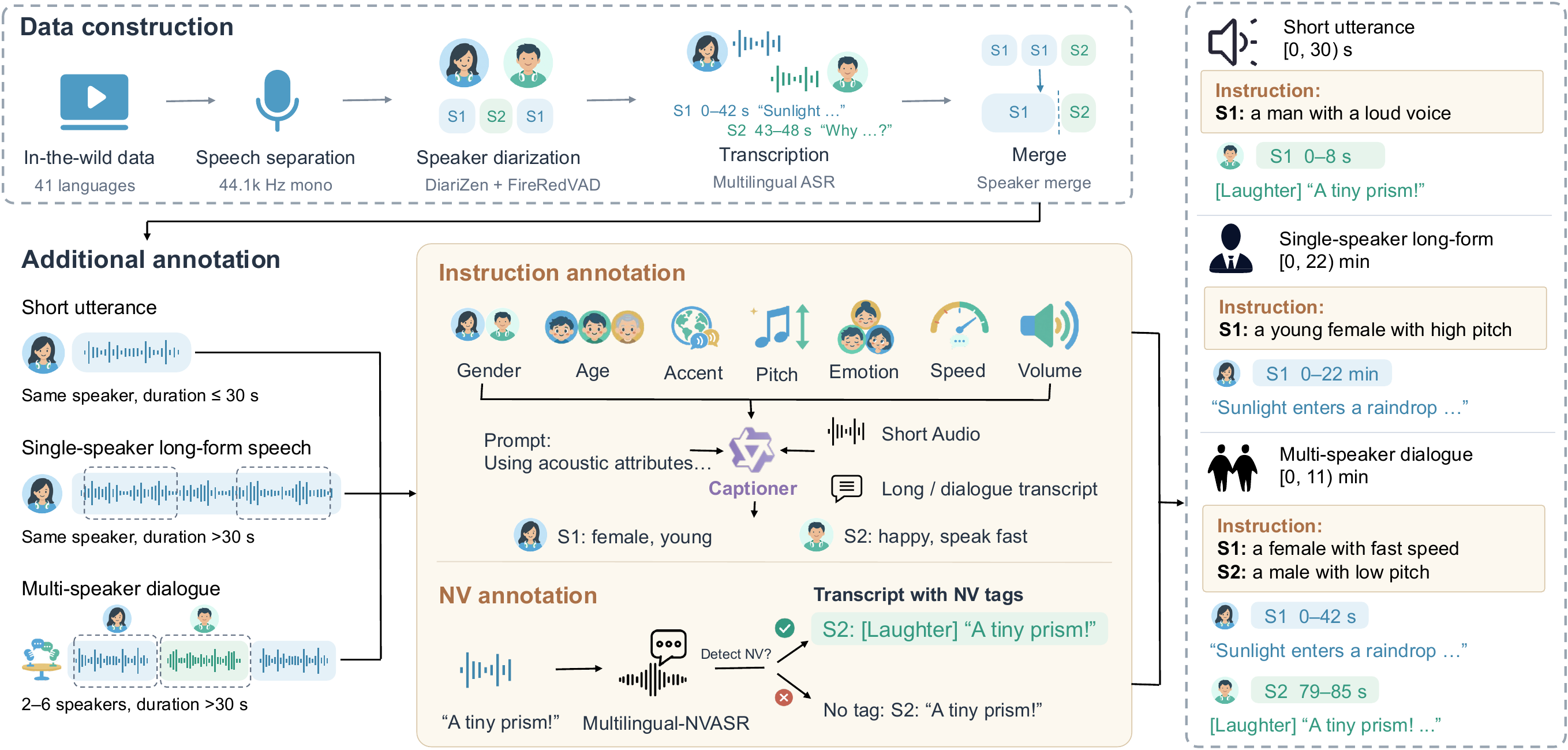}
    \caption{\textbf{Pipeline overview of \emiliapipe.} In-the-wild recordings are enhanced,
    diarized, transcribed, and merged into single-speaker or
    multi-speaker audio. Segments are annotated with natural language style
    instructions and NV tags, then reassembled with speaker and timing information
    into single-speaker long-form speech and multi-speaker dialogue examples.}
    \vspace{-10pt}
    \label{fig:overview}
    \end{figure}

To close these gaps, we introduce \dataset, a 229K-hour dataset across 41 languages for controllable long-form speech generation, as shown in Table~\ref{tab:comparison}. Its construction pipeline, \emiliapipe, organizes source recordings into three complementary views: short utterances, single-speaker long-form speech, and multi-speaker dialogues, while preserving transcripts, speaker assignments, and the original temporal continuity, as shown in Figure~\ref{fig:overview}. These views extend text--speech alignment to long contexts and speaker transitions: examples longer than five minutes contribute 50.0\% of single-speaker long-form hours and 66.8\% of multi-speaker dialogue hours. To support expressive control, annotated subsets provide natural language instructions describing acoustic attributes, with NV tags interleaved with spoken words. Longer examples further capture how these attributes evolve over time, while dialogue instructions explicitly specify speaker attribution. For evaluation, we introduce \dataset-testset, 23~h of bilingual recordings of up to 30~min with up to four speakers. We validate \dataset under two generation paradigms: a non-autoregressive (NAR) masked generative model, Emilia2-NAR \citep{wang2026mgm,zhu2026omnivoice}, and an autoregressive (AR) flow-matching model, Emilia2-AR. With only a 0.6B backbone, Emilia2-NAR achieves the lowest Chinese/English error rates on TTSD-Eval (5.27\%/9.93\%) among compared systems, including the 8B MOSS-TTSD, while Emilia2-AR obtains InstructTTSEval averages of 76.2/78.3, on par with Qwen3-TTS-12Hz-VD (77.1/77.9). These evaluations demonstrate that \dataset effectively supports long-form speech generation and expressive control.

Our contributions are threefold. \textbf{(1)} We establish \dataset and \emiliapipe, a data foundation for controllable long-form speech generation that unifies short utterances, single-speaker long-form speech, and multi-speaker dialogue, enriched with 4.77M instruction-annotated and 933K NV-annotated examples. \textbf{(2)} We introduce \dataset-testset, an open benchmark of 23~h of human-verified bilingual recordings of up to 30~min with up to four speakers, each speaker paired with reference audio. \textbf{(3)} We validate \dataset using two different architectures, AR and NAR, across multiple testsets, showing that \dataset supports long-form and controllable expressive generation across different modeling paradigms.

\begin{table}[t]
\centering
\caption{\textbf{\dataset versus existing TTS training datasets.}
Long denotes duration longer than 30~s. Dialogue denotes multi-speaker
dialogue speech. Instruction denotes natural language instructions. NV denotes nonverbal vocalizations.
$^{\dagger}$Dialogue context prompts for utterance synthesis.}
\label{tab:comparison}
\small
\setlength{\tabcolsep}{3.5pt}
\renewcommand{\arraystretch}{0.95}

\begin{tabular*}{\linewidth}{@{\tableedgepad\extracolsep{\fill}}lrccccc@{\tableedgepad}}
\toprule
\tableheaderrow
Resource & Hours & Language & Long & Dialogue & Instruction & NV \\
\midrule
Emilia \citep{he2025emilia} & 216,000 & 6 & \no & \no & \no & \no \\
MinutesSpeech \citep{nishimura2025hall} & 40,000 & 1 & \yes & \no & \no & \no \\
OpenDialog \citep{zhu2026zipvoice} & 6,800 & 2 & \no & \yes & \no & \no \\
DNASpeech \citep{cheng2025dnaspeech} & 18 & 1 & \no & \yes$^{\dagger}$ & \yes & \no \\
ParaSpeechCaps \citep{diwan2025scaling} & 2,800 & 1 & \no & \no & \yes & \no \\
CapSpeech \citep{wang2026capspeech} & 33,600 & 1 & \no & \no & \yes & \yes \\
Emilia-NV \citep{liao2026emilia} & 573 & 1 & \no & \no & \no & \yes \\
\midrule
\tableoursrow
\textbf{\dataset{} (this work)} & \textbf{229,000} & \textbf{41} & \yes & \yes & \yes & \yes \\
\bottomrule
\end{tabular*}
\end{table}

\section{Related Work}
\label{sec:related_work}
\vspace{-5pt}
\paragraph{Datasets for speech generation.}
Existing datasets have broadened speech supervision, but remain limited in sequence duration or in the temporal granularity of their annotations.
Emilia retains utterances of at most 30~s \citep{he2025emilia}, and
MinutesSpeech extends single-speaker examples only to 180~s
\citep{nishimura2025hall}. For multi-speaker speech, OpenDialog caps dialogue
segments at 30~s \citep{zhu2026zipvoice}. ParaSpeechCaps and CapSpeech annotate
speaker characteristics and style
\citep{diwan2025scaling,wang2026capspeech}, DNASpeech adds dialogue
context and actions \citep{cheng2025dnaspeech}, and Emilia-NV aligns NV events
with text \citep{liao2026emilia}. However, these annotations are likewise
attached to short segments. Recent efforts have begun to construct monologue and dialogue corpora from in-the-wild audio, but no existing open corpus
provides unified speaker, style, and NV annotations over long-form recordings.

\vspace{-5pt}
\paragraph{Benchmarks for long-form and controllable speech generation.}
TTS benchmarks have likewise expanded beyond short-form content alignment.
Long-TTS-Eval focuses on single-speaker long-form speech synthesis
\citep{wang2026mgm}, TTSD-Eval targets dialogue generation
\citep{zhang2026moss}.
InstructTTSEval evaluates natural language instruction following
\citep{huang2025instructttseval}, and NV-Bench evaluates NV synthesis
\citep{ni2026nv}, both on short utterances. Moreover, existing open long-form
benchmarks cover at most two speakers per recording, and only ZipVoice-Dialog
and TTSD-Eval provide per-speaker reference audio
(Table~\ref{tab:long-dialogue-benchmarks}).

\vspace{-5pt}
\paragraph{Controllable and long-form speech models.}
Speech generation increasingly supports structured instructions and longer outputs. Instruction-based TTS maps natural language descriptions to speaking style \citep{guo2023prompttts,yang2024instructtts,zhou2024voxinstruct,chen2026flexivoice,hu2026qwen3}, but is typically trained and evaluated on short utterances. Borderless Long Speech Synthesis and SwanTale extend instruction-based control to extended generation \citep{song2026borderless,zhang2026swantale}. Long-form speech generation has been explored with diverse paradigms, including hierarchical codec language modeling in HALL-E \citep{nishimura2025hall}, autoregressive token prediction in MOSS-TTSD \citep{zhang2026moss}, autoregressive language models coupled with diffusion in MoonCast and VibeVoice \citep{ju2026mooncast,peng2026vibevoice}, and masked acoustic-token prediction in OmniVoice \citep{zhu2026omnivoice}. These diverse modeling paradigms substantially broaden the design space for long-form speech generation.

\begin{figure}[tb]
    \centering
\begin{minipage}[t]{0.47\textwidth}
    \centering
    \includegraphics[height=1.58in]{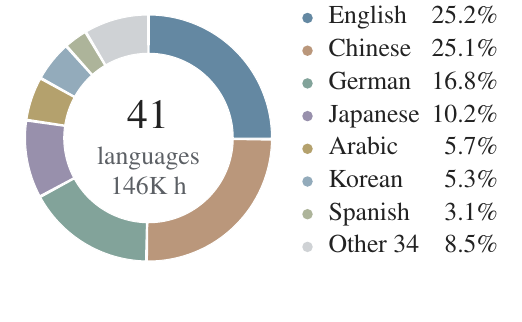}
    \caption{\protect\raggedright\textbf{Language distribution of \shortset.} The 34 remaining languages are grouped as Other.}
    \label{fig:language-distribution}
    \end{minipage}\hfill
    \begin{minipage}[t]{0.51\textwidth}
    \centering
    \includegraphics[height=1.58in]{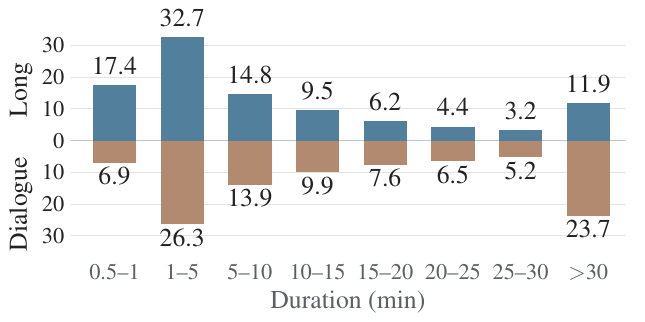}
    \caption{\textbf{Duration distributions of \longset and \dialogueset.} Each subset is normalized separately by audio hours.}
    \label{fig:duration-distribution}
    \end{minipage}
    \end{figure}

\section{\emiliapipe}
\label{sec:pipeline}

\emiliapipe is a processing and annotation pipeline that constructs multilingual
TTS data from in-the-wild recordings. It produces transcribed short utterances,
single-speaker long-form speech, and multi-speaker dialogues, enriched with
style instructions and NV tags.

\subsection{Speech processing and example construction}
\label{subsec:speech-processing}

\paragraph{Audio standardization and separation.}
We inspect source media with FFmpeg, discard recordings sampled below 24~kHz,
and decode the remainder into 44.1~kHz mono float32 audio. Mel-Band RoFormer
\citep{wang2023mel} then extracts the vocal stem, which serves as the audio
for all subsequent processing and for model training.

\vspace{-5pt}
\paragraph{Speaker diarization and segmentation.}
DiariZen \citep{han2025leveraging} identifies speaker-active regions and
builds a speaker timeline for each source recording. Within these regions,
FireRedVAD \citep{xu2026fireredasr2s} refines speech boundaries and splits long
regions into segments for downstream processing. Each segment keeps its source
timestamps and speaker assignment, preserving its correspondence to the
original recording throughout the pipeline.

\vspace{-5pt}
\paragraph{Acoustic quality and synthetic-speech filtering.}
We compute DNSMOS \citep{reddy2022dnsmos} OVR scores to filter low-quality
segments. Given the prevalence of synthetic speech online, an in-house
synthetic-speech classifier flags synthetic Chinese and English speech. Long-form
candidates are assessed as complete examples using aggregated segment-level
scores and flags, rather than requiring every constituent segment to pass the
short-form criteria independently (Appendix~\ref{app:quality-filtering}).

\vspace{-5pt}
\paragraph{Language identification and transcription.}
FireRedLID \citep{xu2026fireredasr2s} identifies the language of each segment
and routes it to an appropriate ASR model: Qwen3-ASR-1.7B \citep{shi2026qwen3},
which supports 30 languages, transcribes those languages, and
IndicConformer-600M \citep{ai4bharat2025indicconformer}, which covers 22 Indic
languages, transcribes the remaining Indic-language segments.

\vspace{-5pt}
\paragraph{Example construction.}
Combining segment-level transcripts with the recording-level speaker timeline,
we construct three views of each recording. For \textbf{short utterances}, we retain utterances of 0.5--30~s
that pass the quality filters above.
For \textbf{single-speaker long-form speech}, we extend across adjacent same-speaker
regions to construct long-form speech of at least 30~s, terminating at speaker changes.
For \textbf{multi-speaker dialogue}, we select windows of at least
30~s containing 2--6 speakers while preserving the original ordering
and timing of speaker turns.

For both long-form views, we retain the continuous vocal-stem audio over each
selected interval, including native pauses and cross-utterance acoustic
transitions, rather than concatenating independently trimmed utterances. Details can be found in Appendix~\ref{app:example-construction}.

\subsection{Instruction and NV annotation}
\label{subsec:expressive}

To support controllable long-form speech generation, we annotate speech with
natural language style instructions and NV tags. To keep annotations accurate,
annotation is limited to segments shorter than five minutes.

\vspace{-5pt}
\paragraph{Instruction annotation.}
We first estimate perceived age and gender, accent, pitch, emotion, speaking
rate, and loudness with task-specific attribute models and signal measurements, as shown in Appendix~\ref{app:expressive}. For segments longer than 30~s, we additionally
retain timestamped intervals that capture changes in emotion, speaking rate,
and loudness. Qwen3-Omni-30B-A3B-Captioner \citep{xu2025qwen3} then
converts these attributes into style instructions. Segments
longer than 30~s use a text-only prompt, following the model card's
recommendation to limit audio inputs to 30~s for fine-grained captioning. Following
InstructTTSEval \citep{huang2025instructttseval}, we design three
format-specific prompts that produce acoustic-parameter specifications
(APS), descriptive-style directives (DSD), and role-play
directions (RP), plus a free-form prompt for a voice description and a
plausible speaker role and scene, as shown in Appendix~\ref{app:instruction-generation}.
All prompts require the instructions to agree with the extracted attributes,
and each dialogue description is tied to its speaker label.

\vspace{-5pt}
\paragraph{Nonverbal annotation.}
We apply Multilingual-NVASR \citep{ni2026nv} to Chinese and English speech,
jointly transcribing spoken words and NV events, such
as \texttt{[Laughter]}, \texttt{[Cough]}, and \texttt{[Sigh]}, into a single interleaved sequence. For
examples containing NV tags, this interleaved sequence serves as the training
transcript, and all other examples retain their ASR transcripts. More details are provided in
Appendix~\ref{app:nv-annotation}.

\begin{table}[tb]
\centering
\caption{\textbf{Scale and annotation coverage of \dataset.}
The three views share source recordings and the Total row sums view entries.
NV counts examples containing NV tags.}
\label{tab:tracks}
\small
\setlength{\tabcolsep}{3pt}
\renewcommand{\arraystretch}{0.95}

\begin{tabular*}{\linewidth}{@{\tableedgepad\extracolsep{\fill}}lrrrrrr@{\tableedgepad}}
\toprule
\tablesubheaderrow
\multirow{2}{*}{View} & \multicolumn{2}{c}{Speech data}
& \multicolumn{2}{c}{Instruction} & \multicolumn{2}{c@{\tableedgepad}}{NV} \\
\cmidrule(lr){2-3}\cmidrule(lr){4-5}\cmidrule(l){6-7}
\tablesubheaderrow
& Examples & Audio (h) & Examples & Audio (h) & Examples & Audio (h) \\
\midrule
\shortset & 99,841,007 & 146,325 & 3,994,566 & 5,102 & 474,078 & 854 \\
\longset & 4,263,913 & 155,688 & 651,060 & 15,317 & 353,929 & 9,572 \\
\dialogueset & 1,538,239 & 96,452 & 120,815 & 4,633 & 104,999 & 4,200 \\
\midrule
\tabletotalrow
\textbf{Total} & \textbf{105,643,159} & \textbf{398,467}
& \textbf{4,766,441} & \textbf{25,053}
& \textbf{933,006} & \textbf{14,626} \\
\bottomrule
\end{tabular*}
\end{table}

\section{\dataset Dataset}
\label{sec:dataset}

Built with \emiliapipe, \dataset comprises short utterances, single-speaker
long-form speech, and multi-speaker dialogues, with subsets enriched by
instruction and NV annotations. Table~\ref{tab:tracks} summarizes the scale and
annotation coverage of each view.

\subsection{\dataset Profile}
\label{subsec:record}

\paragraph{Diverse sources.}
\dataset draws from a broad range of Internet sources, including podcasts and
RSS feeds, livestreams, interviews, public talks, film/TV commentary, sports content, and
games, spanning diverse formats, domains, and speakers. The resulting collection covers 41
languages and approximately 20.5M audio records, 229K hours in total. The source data
are compositionally broad: film/TV commentary accounts for 19.8\% of its hours, followed by humanities (11.5\%),
sports (6.8\%), technology (6.1\%). These
shares illustrate the diversity of source formats, domains, and speakers
represented in the corpus.
Appendix~\ref{app:source-profile} details the source composition.

\vspace{-5pt}
\paragraph{Scale and composition.}
Because the \shortset, \longset, and \dialogueset views are derived from the same source recordings, they partially overlap. The Total row in Table~\ref{tab:tracks} sums the three views (105,643,159 entries and 398,467 hours). After deduplication at the source-audio level, \dataset contains 20,497,038 unique audio records totaling 229,388 hours across 41 languages, including English, Chinese, German, Japanese,
Arabic, Korean, and Spanish, as shown in Figure~\ref{fig:language-distribution}. Ten
languages each contribute more than 1,000~h to \shortset.
More details are provided in Appendix~\ref{app:statistics}.

\vspace{-5pt}
\paragraph{Long-form and expressive coverage.}
Examples longer than five minutes account for 50.0\% of single-speaker
long-form hours and 66.8\% of dialogue hours, as shown in Figure~\ref{fig:duration-distribution}, providing substantial coverage of
extended narration and multi-turn dialogue. Instruction annotations cover
4,766,441 examples, 25,053~hours in total, of which single-speaker long-form speech and
dialogue contribute 19,950~hours (79.6\%), extending instruction supervision well
beyond short utterances. NV tags, annotated for Chinese and English, occur in 933,006 examples spanning 14,626~hours.
Both annotation types are attached to existing examples and add no audio to the
corpus.

\begin{table}[tb]
\centering
\caption{\textbf{Composition of \dataset-testset.}
Hours cover target recordings only. Avg. and Range are the average and range of target recording durations.}
\label{tab:testset-composition}
\small
\setlength{\tabcolsep}{2.2pt}
\renewcommand{\arraystretch}{0.95}
\begin{tabular*}{\linewidth}{@{\tableedgepad\extracolsep{\fill}}lcccccc@{\tableedgepad}}
\toprule
\tablesubheaderrow
Language & Cases & Single-speaker & Dialogue & Audio (h) & Avg. (min) & Range (min) \\
\midrule
Chinese & 59 & 18 & 41 & 11.6 & 11.84 & 0.9--29.3\\
English & 59 & 24 & 35 & 11.7 & 11.86 & 1.0--30.0\\
\midrule
\tabletotalrow
\textbf{Total} & \textbf{118} & \textbf{42} & \textbf{76}
& \textbf{23.3} & \textbf{11.85} & \textbf{0.9--30.0}\\
\bottomrule
\end{tabular*}
\end{table}

\subsection{\dataset-testset}
\label{subsec:testset}

\paragraph{Task and design.}
Existing long-form benchmarks with up to four speakers are either not openly released or do not provide per-speaker reference audio. Among open benchmarks with per-speaker prompts, recordings contain at most two speakers. We therefore introduce \dataset-testset, which, to our knowledge, is the first open long-form benchmark to combine recordings of up to four speakers with reference audio for every speaker. A TTS system must synthesize the entire script as a single recording. As summarized in Table~\ref{tab:testset-composition}, \dataset-testset contains 118 cases totaling 23~hours: 42 single-speaker long-form cases and 76 dialogues with two to four speakers. Target recordings range from 51.1~s to 30.0~min, averaging 11.85~min.

\vspace{-5pt}
\paragraph{Annotation and split protocol.}
Ten annotators verify the target transcripts, speaker assignments, turn boundaries, audio integrity, and the transcript and speaker identity of each prompt. A second annotator independently reviews 10\% of the annotations. Clear errors are corrected directly, and cases with substantial errors are re-annotated. To prevent train--test leakage, isolation is enforced at the level of complete source recordings and duplicate groups: all test sources and their detected duplicates are excluded from the training manifests, so every segment, prompt, and annotation derived from a source belongs to a single split. More details can be found in Appendix~\ref{app:testset}.

\section{Experiments}
\label{sec:experiments}

We evaluate whether \dataset can support long-form generation, multi-speaker synthesis, and expressive control across two different modeling paradigms.

\subsection{Model Architectures}
\label{sec:model}

We study two generative architectures: Emilia2-AR, which combines
autoregressive token prediction with flow-matching acoustic reconstruction, and
Emilia2-NAR, which is trained with masked acoustic-token prediction
(Figure~\ref{fig:model-architectures} in Appendix~\ref{app:model-objectives}).

\vspace{-5pt}
\paragraph{Emilia2-AR.}
\label{subsec:ar-fm}

Emilia2-AR autoregressively predicts a target sequence
$Z=(z_1,\ldots,z_L)$ of speech tokens and speaker-ID markers given conditions
$\mathcal{C}$:
\begin{equation}
p_{\phi}(Z\mid\mathcal{C})
=\prod_{\ell=1}^{L}p_{\phi}(z_{\ell}\mid z_{<\ell},\mathcal{C}),
\label{eq:arfm-factorization}
\end{equation}
where $\mathcal{C}$ comprises the script, instructions, and reference speech. A
flow-matching decoder then generates a mel spectrogram from the predicted
tokens and the selected reference voice, and a vocoder synthesizes the waveform.

\vspace{-5pt}
\paragraph{Emilia2-NAR.}
\label{subsec:nar}

Let $X=(x_{k,t})$ denote the target acoustic-token grid over codebooks
$k\in\{0,\ldots,K-1\}$ and frames $t$, and let a binary mask $M$ produce the
input $\widetilde{X}$, with prompt tokens always visible. Given
$\widetilde{X}$ and the conditions $\mathcal{C}$,
Emilia2-NAR minimizes, for each masked entry,
\begin{equation}
\ell_{k,t}(\theta)
=-\log p_{\theta}(x_{k,t}\mid\widetilde{X},\mathcal{C}),
\qquad M_{k,t}=1.
\label{eq:nar-token-objective}
\end{equation}
Training samples full, coarse-to-fine, and independent per-cell masks with
probabilities 30\%, 50\%, and 20\%. For coarse-to-fine masking, a boundary
$d\sim\mathcal{U}\{0,\ldots,K-1\}$ and
$B_t\sim\mathrm{Bernoulli}(r)$ define
\begin{equation}
M_{k,t}=\begin{cases}
0, & k<d,\\
B_t, & k=d,\\
1, & k>d,
\end{cases}
\qquad r\sim\mathcal{U}(0,1).
\label{eq:nar-depth-mask}
\end{equation}
At inference, the model starts from a fully masked grid and iteratively
decodes tokens in parallel, prioritizing lower-index
codebooks, and an acoustic decoder then reconstructs the waveform.

\subsection{Implementation and Training Details}
\label{subsec:implementation}

\paragraph{Model implementations.}
Both Emilia2 models initialize their sequence-model backbones from Qwen3-0.6B.
Emilia2-AR uses the single-codebook speech tokenizer of CosyVoice~3
\citep{du2025cosyvoice}. Its causal backbone is trained with next-token
prediction over speech tokens and speaker-ID markers. A DiT-based
conditional flow-matching decoder converts the predicted tokens into a mel
spectrogram, then a Pupu-Vocoder
\citep{gu2026aliasing} synthesizes the waveform. Emilia2-NAR adopts
the MGM architecture \citep{wang2026mgm} with the Fish Audio S2 Pro tokenizer
\citep{liao2026fish} for multi-codebook masked-token
prediction, generating outputs in parallel.

\paragraph{Curriculum learning.}
Both models follow the same four-stage curriculum, progressively increasing
context duration and task complexity. Training starts on \shortset with inputs
capped at 30~s, then extends the maximum context to 30~min on \longset. The
third stage trains jointly on \longset and \dialogueset for multi-speaker
dialogue synthesis. Finally, supervised fine-tuning (SFT) on instruction- and
NV-annotated subsets enables expressive control and NV generation. More details can be found in Appendix~\ref{app:training}.

\subsection{Evaluation Setup}
\label{subsec:evaluation-setup}

\paragraph{Testsets.}
We evaluate long-form speech generation on \dataset-testset,
TTSD-Eval \citep{zhang2026moss}, and Long-TTS-Eval \citep{wang2026mgm}, and
short-form instruction following and NV generation on InstructTTSEval
\citep{huang2025instructttseval} and NV-Bench \citep{ni2026nv}, respectively.
For long-form expressive control, we construct two separate sets from
\dataset-testset cases shorter than 5~min: instruction cases use manually
corrected \emiliapipe-generated instructions, while NV cases use DeepSeek-rewritten
scripts with NV tags, details in Appendix~\ref{app:testset-control}.

\vspace{-5pt}
\paragraph{Baselines.}
For long-form generation, we compare with Higgs Audio v2
\citep{higgsaudio2025}, VibeVoice-1.5B \citep{peng2026vibevoice},
and MOSS-TTSD \citep{zhang2026moss}. For short-form instruction following, we
compare with VoxInstruct \citep{zhou2024voxinstruct}, FlexiVoice-Base
\citep{chen2026flexivoice}, and Qwen3-TTS-12Hz-VD \citep{hu2026qwen3}. For
short-form NV generation, we compare with Orpheus-TTS
\citep{canopylabs2025orpheus} (using a fixed preset voice),
SMIIP-NV-CV2 \citep{wu2025smiip}, and CosyVoice~3
\citep{du2025cosyvoice}. Since open long-form baselines are not designed
for explicit expressive control, long-form control evaluations
compare Emilia2-AR and Emilia2-NAR.

\vspace{-5pt}
\paragraph{Objective metrics.}
On \dataset-testset, we report word/character error rates (WER/CER), speaker
similarity (SIM), and speaker-attribution accuracy (ACC), adapting the
TTSD-Eval protocol to recordings of up to 30~min and one to four speakers.
We follow the
official protocols of TTSD-Eval and Long-TTS-Eval, additionally reporting SIM
on Long-TTS-Eval, using the WavLM-based speaker-verification model \citep{anastassiou2024seed}. For
short-form control, InstructTTSEval scores instruction accuracy with
Gemini-2.5-Pro, and NV-Bench uses Multilingual-NVASR to compute PCER/PWER over
NV events and text-only CER/WER. Higher ACC, SIM, and
instruction scores and lower error rates are better. Complete protocols are
given in Appendix~\ref{app:eval-protocol}.

\vspace{-5pt}
\paragraph{Subjective metrics.}
We conduct separate listening tests for the two long-form control tasks. Both
measure speech naturalness (N-MOS), assessed through pronunciation, prosody,
and continuity, independently of control
correctness. Instruction cases measure instruction following capability
(Instr-MOS), and NV cases measure NV Quality, which jointly assesses event
type, placement, and contextual appropriateness. All metrics use a 1--5 scale
and are averaged within each language. More details can be found in
Appendix~\ref{app:listening}.

\begin{table}[tb]
    \centering
    \caption{\textbf{Objective results on \dataset-testset.}
    ACC is accuracy over multi-speaker cases. Single-speaker cases remain
    in SIM and CER/WER. Bold and underlining
    mark the best and second-best per column.}
    \label{tab:emilia2-testset-results}
    \small
    \setlength{\tabcolsep}{2pt}
    \renewcommand{\arraystretch}{0.95}
    \begin{tabular*}{\linewidth}{@{\tableedgepad\extracolsep{\fill}}lc*{6}{c}@{\tableedgepad}}
    \toprule
    \tableheaderrow
    \multirow{2}{*}{Model} & \multirow{2}{*}{\shortstack{Params.}}
    & \multicolumn{3}{c}{Chinese} & \multicolumn{3}{c@{\tableedgepad}}{English} \\
    \cmidrule(lr){3-5}\cmidrule(l){6-8}
    \tablesubheaderrow
    & & ACC$\uparrow$ & SIM$\uparrow$ & CER$\downarrow$
    & ACC$\uparrow$ & SIM$\uparrow$ & WER$\downarrow$ \\
    \midrule
    \tabletotalrow GT & -- & 99.33 & 0.88 & 3.00 & 98.38 & 0.85 & 3.00 \\
    \midrule
    Higgs Audio v2 & 3B & 75.00 & 0.74 & 75.46 & 72.84 & 0.67 & 81.38 \\
    VibeVoice-1.5B & 1.5B & 93.78 & 0.81 & 4.61 & 95.98 & 0.75 & \textbf{5.66} \\
    MOSS-TTSD & 8B & \textbf{99.28} & \textbf{0.87} & \textbf{2.18} & \textbf{99.22} & \textbf{0.81} & \underline{5.83} \\
\midrule
    \tableoursrow \textbf{Emilia2-AR} & 0.6B & \underline{98.57} & \underline{0.84} & \underline{3.67} & 95.05 & 0.74 & 6.82 \\
    \tableoursrow \textbf{Emilia2-NAR} & 0.6B & 97.38 & \underline{0.84} & 8.62 & \underline{96.46} & \underline{0.79} & 6.62 \\
    \bottomrule
    \end{tabular*}
    \end{table}

\subsection{Main Results}
\label{subsec:results}

Tables~\ref{tab:emilia2-testset-results} and~\ref{tab:main-benchmark-results}
compare Emilia2 with existing mainstream long-form speech generation systems. Despite using only a 0.6B backbone, both Emilia2 variants outperform the 3B Higgs Audio v2 across all reported evaluation measures, while Emilia2-NAR surpasses VibeVoice-1.5B on all reported results in TTSD-Eval and Long-TTS-Eval. Both variants are also competitive with the substantially larger 8B MOSS-TTSD. On TTSD-Eval, Emilia2-NAR achieves the lowest Chinese CER and English WER, while nearly matching MOSS-TTSD in speaker attribution accuracy: its ACC reaches 95.30\% in Chinese and 94.89\% in English, compared with 95.36\% and 95.00\% for MOSS-TTSD, respectively. On Long-TTS-Eval, Emilia2-AR achieves the highest SIM in both languages. On \dataset-testset, an Emilia2 variant ranks second only to MOSS-TTSD on five of the six reported results. Overall, the two generation paradigms exhibit complementary strengths: Emilia2-NAR leads Emilia2-AR on every TTSD-Eval metric and on Long-TTS-Eval error rates, whereas Emilia2-AR attains the highest SIM on single-speaker Long-TTS-Eval. On \dataset-testset, Emilia2-AR performs better in Chinese and Emilia2-NAR in English.

Both models further demonstrate strong instruction controllability. On InstructTTSEval, as shown in Table~\ref{tab:instruction-control-results}, Emilia2-AR is competitive with the state-of-the-art Qwen3-TTS-12Hz-VD and outperforms it on four of the six criteria. On NV-Bench, as shown in Table~\ref{tab:nv-control-results}, Emilia2-NAR achieves the lowest PCER/PWER and text-only CER/WER, together with the highest SIM, across all subsets and both languages. The minute-scale long-form subjective evaluations reveal a similar complementarity: Emilia2-AR scores slightly higher in N-MOS and Instr-MOS, whereas Emilia2-NAR delivers higher NV Quality and lower content error under NV conditioning.

\begin{table}[tb]
\centering
\caption{\textbf{Long-form synthesis on TTSD-Eval and Long-TTS-Eval.} ACC denotes speaker-attribution accuracy, SIM speaker similarity, and CER/WER character/word error rate. Bold and
underlining mark the best and second-best scores per column.}
\label{tab:main-benchmark-results}
\footnotesize
\setlength{\tabcolsep}{2pt}
\renewcommand{\arraystretch}{0.95}
\begin{tabular*}{\linewidth}{@{\tableedgepad\extracolsep{\fill}}l*{10}{r}@{\tableedgepad}}
\toprule
\tableheaderrow
\multirow{3}{*}{Model}
& \multicolumn{6}{c}{TTSD-Eval} & \multicolumn{4}{c@{\tableedgepad}}{Long-TTS-Eval} \\
\cmidrule(lr){2-7}\cmidrule(l){8-11}
\tablesubheaderrow
& \multicolumn{3}{c}{Chinese} & \multicolumn{3}{c}{English}
& \multicolumn{2}{c}{Chinese} & \multicolumn{2}{c@{\tableedgepad}}{English} \\
\cmidrule(lr){2-4}\cmidrule(lr){5-7}\cmidrule(lr){8-9}\cmidrule(l){10-11}
\tablesubheaderrow
& ACC$\uparrow$ & SIM$\uparrow$ & CER$\downarrow$
& ACC$\uparrow$ & SIM$\uparrow$ & WER$\downarrow$
& CER$\downarrow$ & SIM$\uparrow$ & WER$\downarrow$ & SIM$\uparrow$ \\
\midrule
Higgs Audio v2 & 60.97 & 0.63 & 79.32 & 70.60 & 0.62 & 47.54 & 73.65 & 0.60  & 73.30 & 0.47  \\
VibeVoice-1.5B & 88.25 & 0.73 & 9.21 & 93.58 & 0.68 & \underline{10.18} & 16.98 & 0.67 & 11.20 & 0.59 \\
MOSS-TTSD & \textbf{95.36} & \textbf{0.80} & \underline{6.08} & \textbf{95.00} & \textbf{0.74} & 11.09 & \textbf{8.22} & \underline{0.74} & \textbf{8.54} & \underline{0.63} \\
\midrule
\tableoursrow \textbf{Emilia2-AR} & 92.89 & 0.75 & 6.92 & 92.66 & 0.65 & 12.25 & 14.42 & \textbf{0.75} & 17.76 & \textbf{0.67} \\
\tableoursrow \textbf{Emilia2-NAR} & \underline{95.30} & \underline{0.78} & \textbf{5.27} & \underline{94.89} & \underline{0.73} & \textbf{9.93} & \underline{13.05} & \underline{0.74} & \underline{9.79} & 0.62 \\
\bottomrule
\end{tabular*}
\end{table}

\begin{table}[tb]
    \centering
    \caption{\textbf{Instruction following at two timescales.}
    Short-form results are on InstructTTSEval, and long-form results on
    instruction-conditioned \dataset-testset cases. In the
    short-form block, bold and underlining mark the best and second-best
    scores per column.}
    \label{tab:instruction-control-results}
    {\small
    \setlength{\tabcolsep}{1.2pt}
    \renewcommand{\arraystretch}{0.95}
    \begin{tabular*}{\linewidth}{@{\tableedgepad\extracolsep{\fill}}l*{8}{c}@{\tableedgepad}}
    \toprule
    \tableheaderrow
    \emph{Short-form} & \multicolumn{4}{c}{Chinese}
    & \multicolumn{4}{c@{\tableedgepad}}{English} \\
    \cmidrule(lr){2-5}\cmidrule(l){6-9}
    \tablesubheaderrow
    Model & APS & DSD & RP & Avg. & APS & DSD & RP & Avg. \\
    \midrule
    VoxInstruct & 47.5 & 52.3 & 42.6 & 47.5 & 54.9 & 57.0 & 39.3 & 50.4 \\
    FlexiVoice-Base & 56.7 & 59.1 & 59.5 & 58.4 & 63.6 & 75.0 & 60.6 & 66.4 \\
    Qwen3-TTS-12Hz-VD & \textbf{85.2} & \textbf{81.1} & \underline{65.1} & \textbf{77.1} & \underline{82.9} & \underline{82.4} & \underline{68.4} & \underline{77.9} \\
    \midrule
    \tableoursrow \textbf{Emilia2-AR} & \underline{82.0} & \underline{72.5} & \textbf{74.0} & \underline{76.2} & \textbf{83.0} & \textbf{83.5} & \textbf{68.5} & \textbf{78.3} \\
    \tableoursrow \textbf{Emilia2-NAR} & 69.2 & 59.8 & 55.6 & 61.6 & 67.4 & 66.5 & 49.8 & 61.2 \\
    \bottomrule
    \end{tabular*}
    \begin{tabular*}{\linewidth}{@{\tableedgepad\extracolsep{\fill}}l*{6}{c}@{\tableedgepad}}
    \tableheaderrow
    \emph{Long-form} & \multicolumn{3}{c}{Chinese}
    & \multicolumn{3}{c@{\tableedgepad}}{English} \\
    \cmidrule(lr){2-4}\cmidrule(l){5-7}
    \tablesubheaderrow
    Model & CER$\downarrow$ & N-MOS$\uparrow$ & Instr-MOS$\uparrow$
    & WER$\downarrow$ & N-MOS$\uparrow$ & Instr-MOS$\uparrow$ \\
    \midrule
    \tableoursrow \textbf{Emilia2-AR} & 4.87 & 4.40 $\pm$ 0.30 & 3.56 $\pm$ 0.40
    & 2.60 & 4.32 $\pm$ 0.20 & 3.68 $\pm$ 0.38 \\
    \tableoursrow \textbf{Emilia2-NAR} & 3.21 & 4.20 $\pm$ 0.20 & 3.40 $\pm$ 0.38
    & 4.38 & 4.28 $\pm$ 0.24 & 3.52 $\pm$ 0.38 \\
    \bottomrule
    \end{tabular*}
    }
    \end{table}
    
    \begin{table}[t]
    \centering
    \caption{\textbf{NV generation at two timescales.} Short-form results
    are on NV-Bench, and long-form results on NV-conditioned \dataset-testset
    cases. In the short-form block, bold and underlining
    mark the best and second-best scores per column.}
    \label{tab:nv-control-results}
    {\small
    \setlength{\tabcolsep}{1.2pt}
    \renewcommand{\arraystretch}{0.95}
    \begin{tabular*}{\linewidth}{@{\tableedgepad\extracolsep{\fill}}l*{12}{c}@{\tableedgepad}}
    \toprule
    \tableheaderrow
    \emph{Short-form}
    & \multicolumn{6}{c}{Single-label subset}
    & \multicolumn{6}{c@{\tableedgepad}}{Multi-label subset} \\
    \cmidrule(lr){2-7}\cmidrule(l){8-13}
    \tablesubheaderrow
    \multirow{2}{*}{Model}
    & \multicolumn{3}{c}{Chinese} & \multicolumn{3}{c}{English}
    & \multicolumn{3}{c}{Chinese} & \multicolumn{3}{c@{\tableedgepad}}{English} \\
    \cmidrule(lr){2-4}\cmidrule(lr){5-7}\cmidrule(lr){8-10}\cmidrule(l){11-13}
    \tablesubheaderrow
    & PCER$\downarrow$ & CER$\downarrow$ & SIM$\uparrow$
    & PWER$\downarrow$ & WER$\downarrow$ & SIM$\uparrow$
    & PCER$\downarrow$ & CER$\downarrow$ & SIM$\uparrow$
    & PWER$\downarrow$ & WER$\downarrow$ & SIM$\uparrow$ \\
    \midrule
    Orpheus-TTS
    & 88.77 & 11.36 & -- & 71.92 & 9.03 & --
    & 84.85 & 19.83 & -- & 71.46 & 8.68 & -- \\
    SMIIP-NV-CV2
    & 75.64 & 8.80 & 0.72 & 56.80 & 17.92 & 0.58
    & 77.20 & 10.66 & \underline{0.72} & \underline{54.49} & 20.93 & 0.58 \\
    CosyVoice 3
    & 57.69 & \underline{3.85} & \underline{0.76} & 62.75 & \underline{7.87} & \underline{0.70}
    & 61.94 & \underline{4.75} & \underline{0.72} & 57.84 & \underline{6.39} & \underline{0.72} \\
    \midrule
    \tableoursrow \textbf{Emilia2-AR}
    & \underline{39.23} & 5.39 & 0.67 & \underline{55.59} & 10.96 & 0.62
    & \underline{49.48} & 6.25 & 0.67 & 57.19 & 9.10 & 0.62 \\
    \tableoursrow \textbf{Emilia2-NAR}
    & \textbf{28.15} & \textbf{2.36} & \textbf{0.78} & \textbf{40.12} & \textbf{6.21} & \textbf{0.73}
    & \textbf{38.07} & \textbf{3.87} & \textbf{0.78} & \textbf{41.77} & \textbf{4.85} & \textbf{0.76} \\
    \bottomrule
    \end{tabular*}
    \begin{tabular*}{\linewidth}{@{\tableedgepad\extracolsep{\fill}}l*{6}{c}@{\tableedgepad}}
    \tableheaderrow
    \emph{Long-form} & \multicolumn{3}{c}{Chinese}
    & \multicolumn{3}{c@{\tableedgepad}}{English} \\
    \cmidrule(lr){2-4}\cmidrule(l){5-7}
    \tablesubheaderrow
    Model & CER$\downarrow$ & N-MOS$\uparrow$ & \shortstack{NV Quality$\uparrow$}
    & WER$\downarrow$ & N-MOS$\uparrow$ & \shortstack{NV Quality$\uparrow$} \\
    \midrule
    \tableoursrow \textbf{Emilia2-AR} & 5.44 & 4.44 $\pm$ 0.24 & 3.40 $\pm$ 0.40
    & 3.82 & 4.48 $\pm$ 0.24 & 3.40 $\pm$ 0.36 \\
    \tableoursrow \textbf{Emilia2-NAR} & 2.58 & 4.28 $\pm$ 0.20 & 3.76 $\pm$ 0.32
    & 3.26 & 4.40 $\pm$ 0.26 & 3.88 $\pm$ 0.46 \\
    \bottomrule
    \end{tabular*}
    }
    \end{table}

\subsection{Analysis}
\label{subsec:analysis}

\paragraph{\dataset supports both long-form generation and expressive control.}
Trained on \dataset, the 0.6B-backbone Emilia2-AR and Emilia2-NAR models demonstrate both capabilities across AR and NAR architectures. In long-form generation, both remain competitive with the 8B MOSS-TTSD despite having roughly one-thirteenth as many backbone parameters. For expressive control, Emilia2-AR matches Qwen3-TTS-12Hz-VD on InstructTTSEval, while Emilia2-NAR achieves the lowest PCER/PWER on every NV-Bench subset. On minute-scale instruction- and NV-conditioned speech, both models achieve N-MOS $\geq$ 4.20, Instr-MOS $\geq$ 3.40, and NV Quality $\geq$ 3.40 in both languages. Together, these results show that \dataset supports both extended generation and explicit expressive control with high naturalness.

\noindent
\begin{minipage}[t]{0.475\linewidth}
\vspace{0pt}
\paragraph{\dataset-testset exposes failure modes hidden by aggregate scores.}
By stratifying performance by speaker count and recording duration, \dataset-testset probes conditions that existing benchmarks cover only sparsely (Figure~\ref{fig:testset-chinese-cer-stratified}). GT CER stays low across strata, so the trends reflect generation errors rather than ASR artifacts. Emilia2-AR and the much larger MOSS-TTSD show no systematic deterioration along either axis. VibeVoice, however, becomes less accurate as speaker count grows, a weakness masked by its competitive aggregate CER. Emilia2-NAR degrades as recording duration increases, consistent with its fixed decoding-step budget. Although chunking reduces Emilia2-NAR’s Chinese CER from 8.62\% to 1.73\%, it lowers N-MOS in all four model–language pairs, by up to 0.75 (Appendix~\ref{app:chunk-vs-whole}). This contrast shows that long-form synthesis cannot be reduced to independently generated short chunks: CER/WER captures local content accuracy but misses cross-chunk prosodic continuity.
\end{minipage}\hfill
\begin{minipage}[t]{0.5\linewidth}
\vspace{0pt}
\centering
\includegraphics[width=\linewidth]{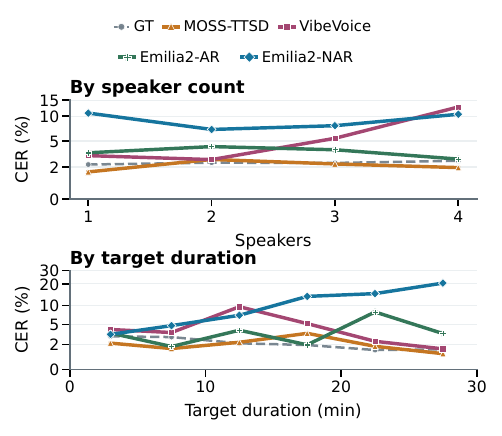}
\captionof{figure}{\protect\raggedright\textbf{Stratified Chinese CER on \dataset-testset.}
Top: by speaker count, bottom: by target duration.}
\label{fig:testset-chinese-cer-stratified}
\end{minipage}

\section{Conclusion}
\label{sec:conclusion}

We introduced \dataset, a 229K-hour, 41-language dataset for controllable
long-form speech generation. Built with \emiliapipe, it unifies short utterances,
single-speaker long-form speech, and multi-speaker dialogue with instruction and
NV annotations. \dataset-testset adds 23 hours of human-verified bilingual
recordings of up to 30~min with up to four speakers. Experiments with 0.6B Emilia2-AR and
Emilia2-NAR models show that \dataset supports both long-form generation and expressive
control, with complementary strengths: Emilia2-NAR excels in dialogue content
alignment and NV Quality, and Emilia2-AR in single-speaker speaker similarity and
naturalness. These results demonstrate that \dataset can serve as an open
data foundation for long-form speech generation and minute-scale expressive control.

\subsection*{AI use statement}

We used generative AI tools for dataset annotation, evaluation-set
construction, and manuscript editing. As described in
Section~\ref{subsec:expressive} and Appendix~\ref{app:instruction-generation},
Qwen3-Omni-30B-A3B-Captioner is a component of the dataset construction
pipeline that generates natural language instructions from extracted
acoustic attributes and, where applicable, audio or transcript context. As
described in Appendix~\ref{app:testset-control}, DeepSeek was used to rewrite
scripts and insert case-specific nonverbal-vocalization tags for the
NV-conditioned evaluation set. Gemini-2.5-Pro serves as the automatic judge in
the official InstructTTSEval protocol. Generative AI tools were also used to
improve the language and clarity of the manuscript. The authors reviewed all
AI-assisted outputs and take full responsibility for the content of this paper.

\subsection*{Ethics statement}
\label{sec:ethics}

Speech can reveal identity and sensitive personal information, and voice
generation can enable impersonation. Our source review therefore separately
considers permission to process source audio, to use it for model training, and
to publish or redistribute derived artifacts. Public accessibility alone is not
treated as consent or a redistribution right. The planned public release
consists of source-linked URL manifests and distributable derived artifacts,
including annotations, metadata, and processing configurations, together with
the preprocessing and evaluation code needed for reproducibility. Stable
identifiers will propagate corrections and removal requests across hosted views
and later manifests, and data will document annotation
licenses and residual risks. Age and gender annotations are perceived,
model-derived attributes rather than speaker metadata.

\subsection*{Reproducibility statement}
\label{sec:reproducibility}

We document the full construction pipeline and filtering thresholds
(Section~\ref{sec:pipeline} and Appendix~\ref{app:construction-annotation}),
component validation (Appendix~\ref{app:validation}), the testset protocol
(Appendix~\ref{app:testset}), and the model, training, and evaluation settings
(Section~\ref{sec:experiments} and Appendices~\ref{app:training}
and~\ref{app:eval-protocol}). The release includes versioned
manifests, processing configurations, annotation provenance, and model training
and inference configurations. Dataset statistics are tied to frozen manifest
checksums and recorded source-access dates (Appendix~\ref{app:release-details}).

\bibliography{references}
\bibliographystyle{plainnat}

\appendix
\raggedbottom

\clearpage

\section{Dataset Construction and Expressive Annotation}
\label{app:construction-annotation}

This appendix details the procedures summarized in Section~\ref{sec:pipeline}:
the filtering and construction rules for the three speech views
(Appendix~\ref{app:example-construction}), acoustic attribute extraction
(Appendix~\ref{app:expressive}), instruction generation
(Appendix~\ref{app:instruction-generation}), and NV annotation
(Appendix~\ref{app:nv-annotation}). Validation of the main automatic components
is reported separately in Appendix~\ref{app:validation}.

\subsection{Filtering and Example Construction}
\label{app:example-construction}

\paragraph{Acoustic quality and synthetic-speech filtering.}
\label{app:quality-filtering}
Short-form selection requires a DNSMOS overall quality (OVR) score of at
least 3.0 for Chinese and English and 2.4 for other languages
\citep{he2025emilia}, plus a check by the in-house synthetic-speech classifier
for Chinese and English.
Long-form passages and dialogues are assessed as complete examples:
their duration-weighted mean OVR must reach 2.4, and a Chinese or English
example fails the synthetic-speech check only when more than 50\% of
its duration is flagged.

\paragraph{Short utterances.}
Adjacent same-speaker segments separated by less than 3~s are merged before
filtering. We retain the resulting segments when they are 0.5--30~s long and
pass the acoustic-quality and applicable synthetic-speech checks above.
Each accepted segment must have a supported language, a language-identification
confidence of at least 0.8, and a valid transcript containing at least two
characters. We further require the ratio of voiced
duration to transcript length to lie within $[0.015, 1.0]$~s per character for
alphabetic scripts and $[0.04, 2.0]$~s per character for Chinese, Japanese,
Korean, and Thai. These examples provide the local text--audio pairs used for
short-form synthesis.

\paragraph{Single-speaker long-form speech.}
Adjacent segments assigned to the same speaker are merged within the original
recording, bridging pauses of up to 6~s and stopping at a speaker change. We
retain passages of at least 30~s, with no upper length bound, that satisfy the
long-form filtering criteria above. Each accepted passage preserves the original
pauses and acoustic transitions across sentences, providing a complete
long-form training target.

\paragraph{Multi-speaker dialogue.}
We divide each recording at silences longer than 6~s and select windows
of at least 30~s containing two to six speakers. Each window must contain at
least $\max(2, N_{\text{spk}})$ turns, with at least 15\% of the turns assigned
to the second-most-active speaker. Windows must satisfy the same long-form
filtering criteria and are discarded if overlapping speech from two or more
speakers covers more than 5\% of their duration. Beforehand, diarized regions
whose overlap with another speaker exceeds 2~s or 40\% of their duration are
removed. Retained dialogues preserve the original order and timing of
exchanges, with each turn linked to its speaker and transcript.

\paragraph{Export and shared source audio.}
The exporter writes the separated vocal stem of each source recording to one
M4A file. Short utterances,
single-speaker long-form speech, and multi-speaker dialogue windows are stored as start--end intervals in
their manifests, each pointing to that file. This avoids storing the same
audio repeatedly and allows a short utterance to be located within its
longer context. Overlapping intervals count toward multiple views but only
once when measuring their combined audio coverage.

\subsection{Acoustic Attribute Extraction}
\label{app:expressive}

\paragraph{Voice attributes.}
We use
wav2vec2-large-robust-24-ft-age-gender\footnote{\href{https://huggingface.co/audeering/wav2vec2-large-robust-24-ft-age-gender}{\nolinkurl{https://huggingface.co/audeering/}}\\
\href{https://huggingface.co/audeering/wav2vec2-large-robust-24-ft-age-gender}{\nolinkurl{wav2vec2-large-robust-24-ft-age-gender}}}
for a three-class female/male/child prediction and one continuous age estimate.
Wavlm-large-age-sex\footnote{\url{https://huggingface.co/tiantiaf/wavlm-large-age-sex}}
provides a second age estimate. The age tag is child when the first model
predicts child. Otherwise, we bin the median of the two age estimates.
Both continuous estimates are retained in the annotations.
Chinese accent labels use a FireRedLID$\rightarrow$Voxlect \citep{feng2026voxlect} cascade.
FireRedLID supplies Wu, Min, and Xiang, which the Voxlect Mandarin head does
not cover, but collapses northern Mandarin subgroups into one north class, so
Voxlect then refines north and other routes into the supported Mandarin
subgroups and also labels Cantonese. Separate Voxlect heads cover English
and Arabic accents, and PENN\footnote{\url{https://github.com/interactiveaudiolab/penn}}
estimates pitch. These attributes describe the
perceived voice and serve as acoustic evidence for instruction generation. Age
and gender are treated as perceived attributes rather than speaker metadata.

\paragraph{Style attributes and temporal changes.}
An in-house speech emotion recognizer predicts seven classes: neutral, happy,
angry, sad, surprised, fearful, and disgusted. We retain a prediction only
when its confidence reaches 0.7. Lower-confidence cases remain unlabeled
rather than being treated as neutral.
Speaking rate is measured in characters per second (CPS), and
integrated loudness is measured in LUFS. Rate values are interpreted
according to the language and transcript representation.
For segments longer than 30~s, we retain intervals describing changes in emotion,
speaking rate, and loudness, together with their measurements and available
confidence scores. Each interval specifies both the style attribute
and the part of the recording to which it applies.
These intervals capture within-recording changes that a single global style
label would miss.
Component selection for all attribute extractors is reported in
Appendix~\ref{app:attribute-validation}.

\subsection{Instruction Generation}
\label{app:instruction-generation}

\paragraph{Captioner inputs.}
We use Qwen3-Omni-30B-A3B-Captioner to generate natural language style
instructions from the available acoustic evidence. For segments of at most
30~s, the captioner receives the waveform and a text prompt containing the
extracted voice and style attributes. Longer segments use a text-only
prompt, following the model card's recommendation to limit audio inputs to
30~s for fine-grained captioning. All four forms
receive the extracted attributes. Only the free-form prompt also receives the
transcript, which is used for broad register and scene context rather than
acoustic inference. Missing attributes are omitted. For segments longer than 30~s, a
tag line describes measured changes in loudness, speaking rate, and emotion
chronologically with their positions in the recording.

\paragraph{Instruction forms and content constraints.}
Following the prompt design of InstructTTSEval \citep{huang2025instructttseval},
we use three format-specific prompts. \emph{Acoustic-parameter
specifications} (APS) express voice attributes as compact, semicolon-separated
lists. \emph{Descriptive-style directives} (DSD) describe pitch, pace,
loudness, and emotional character in natural prose. \emph{Role-play
directions} (RP) evoke an abstract role, mood, or atmosphere while remaining
consistent with the supplied attributes. A fourth, free-form prompt
imposes no output-format constraint and asks for a voice description and a
plausible, general speaker role and scene. APS, DSD, RP, and free-form
instructions must agree
with the supplied attributes, and must not quote, paraphrase, or retell the
spoken content. RP imagery is synthetic performance guidance rather than a
factual account of the recording context.

\paragraph{Language and dialogue organization.}
Instruction form and output language are selected deterministically for each
example so that annotation choices remain reproducible. English and Chinese
instruction versions are supported independently of the audio language.
Dialogue prompts repeat the tag block per speaker and request one independently
grounded \texttt{S1:}, \texttt{S2:}, etc.\ line in speaker order. Only free form
receives the speaker-marked transcript.

\paragraph{Captioner prompt templates.}
The following gives the task clauses of the English prompt templates, with
\texttt{[TAGS]} and \texttt{[TRANSCRIPT]} standing for per-record
substitutions. The tag block contains available gender, perceived
age band, accent, pitch, volume, speaking rate, and gated emotion labels.
Prompts for segments longer than 30~s can add a timestamped line for within-recording
changes. The age field is a phrase such as ``a young adult,'' not a regressed
number of years. In text-only mode, \texttt{[TRANSCRIPT]} is inserted only
for free-form instructions and is truncated to 3,000 characters when needed.
APS and DSD include fixed format examples. RP rotates among six examples by
record ID. The free-form example is disabled in the corpus run.

\Needspace{15\baselineskip}
\begin{promptbox}{Free form}
``Your task is to generate ONE caption that (a)
describes the characteristics of the speaker's voice and (b) infers a
plausible identity and scene for the speaker. Use the following tags EXACTLY
as given---describe each with the labeled value, do NOT replace it with your
own impression from the audio: \texttt{[TAGS]} \texttt{[TRANSCRIPT]} Then, in
the same caption, briefly infer WHO the speaker is likely to be (a plausible
role or persona) and WHAT scene or situation they seem to be in, consistent
with the voice.''\par\smallskip
\textbf{Constraints.} The attached critical rules forbid quoting or retelling
speech content, replacing a tag with the model's own acoustic impression,
describing background sounds or recording quality, and inventing a specific
scene from the spoken words.\par\smallskip
\textbf{Output.} \texttt{YOUR CAPTION:}
\end{promptbox}

\begin{promptbox}{APS: acoustic-parameter specification}
``Task: produce an APS (acoustic-parameter spec)---a
terse, semicolon-separated list of voice attributes only. No sentences, no
scene, no identity, no speech content. Use these tags exactly (do not alter
them): \texttt{[TAGS]}''\par\smallskip
\textbf{Output.} After the rules and format example: \texttt{YOUR APS:}
\end{promptbox}

\begin{promptbox}{DSD: descriptive-style directive}
``Task: produce a DSD (descriptive-style directive)---
flowing prose describing pitch, pace, volume, emotional undertone, and general
character/temperament. Use these tags exactly: \texttt{[TAGS]}''\par\smallskip
\textbf{Output.} After the rules and format example: \texttt{YOUR DSD:}
\end{promptbox}

\begin{promptbox}{RP: role-play direction}
``Task: produce an RP (role-play) voice direction---an
evocative prompt that casts the speaker as an abstract ROLE within a MOOD or
ATMOSPHERE, inferred from how the VOICE SOUNDS (pitch, pace, volume, energy,
timbre) and NOT from what is being said. Lean on imagery rather than a list of
acoustic parameters. VARY your structure: open on the scene, the character,
a sensory image, or a direct instruction---do NOT fall into a fixed formula
or a stock opening. A dynamic arc that shifts tone is welcome. Tags (a light
guide---weave them into the imagery, don't just list them; anything you state
must match): \texttt{[TAGS]}''\par\smallskip
\textbf{Output.} After the rules and rotated example: \texttt{YOUR RP:}
\end{promptbox}

\begin{promptbox}{Shared rules and examples}
\textbf{Common guard.} APS and DSD append the following constraints: ``Never quote, paraphrase, or
retell what is being said---not the words, events, actions, names, or plot;
and never use quotation marks. Judge the voice by its SOUND (pitch, pace,
volume, energy, timbre), NOT by the meaning of the words---as if you did not
understand the language. Describe only the human voice (not background sounds
or audio quality). State the volume exactly as the tag says (loud / moderate
/ soft)---an energetic or fast delivery is not the same as loud volume.''
Free form additionally requires a plausible, general role and scene rather
than specific content inferred from the words. In text-only mode, its content
rule instead says the transcript may inform only the general register and
setting; acoustic claims must come from tags. RP uses a separate rule that
forbids deriving its role or scene from topic, plot, events, or meaning and
requires every concrete attribute it mentions to match its tag. All forms
describe standard Mandarin without calling it an accent.

\par\smallskip\textbf{Format examples.} The APS format example is ``Male, 40--55 years old; low-pitched, deep timbre;
measured pace, moderate volume; standard Mandarin.'' The DSD example is ``A
deep, resonant voice at a medium pitch, delivered at a steady, measured pace
and a moderate volume, with a calm and composed emotional undertone that
conveys a mature, steady, and responsible character.'' RP rotates one of six
role-and-scene examples; each prompt instructs the captioner to invent its own
imagery rather than copy the example. Chinese prompts use corresponding Chinese
task and constraint text.
\end{promptbox}

\Needspace{8\baselineskip}
\subsection{Nonverbal Vocalization Annotation}
\label{app:nv-annotation}

We annotate NV content with Multilingual-NVASR, released with NV-Bench
\citep{ni2026nv}. Built on SenseVoice-Small, the recognizer supports Chinese and
English and transcribes spoken words together with structured tags such as
\texttt{[Laughter]}, \texttt{[Cough]}, and \texttt{[Sigh]}, marking each NV
event and its order relative to the surrounding words. For examples containing
at least one NV tag, the NVASR output, with its interleaved tags, replaces the
Qwen3-ASR transcript as the training text. Examples without NV tags retain
their original ASR transcripts. The two annotation types play
complementary roles: instructions describe how to perform a passage, while NV
tags specify discrete acoustic events within it.

\section{Quality Control and Component Validation}
\label{app:validation}

Quality control in \emiliapipe operates at complementary levels. Automatic
acoustic, duration, overlap, transcript-density, and synthetic-speech rules
are applied at corpus scale to decide which examples enter the three training
views (Appendix~\ref{app:example-construction}). Each automatic component is
selected on public benchmarks and in-house diagnostics: language identification
(Appendix~\ref{app:lid-validation}), transcription
(Appendix~\ref{app:asr-validation}), speech segmentation
(Appendix~\ref{app:segmentation-validation}), and expressive-attribute
extraction (Appendix~\ref{app:attribute-validation}). The extracted
measurements are retained as provenance for every generated instruction, and
the evaluation benchmark is fully verified by human annotators
(Appendix~\ref{app:testset}).

\subsection{Language Identification}
\label{app:lid-validation}

\paragraph{Published benchmarks.}
Table~\ref{tab:fireredlid-published} collects representative published results
for FireRedLID, the language identification (LID) component of \emiliapipe.
FireRedLID is compared with the strongest baseline reported for each benchmark
and performs well across multilingual and dialect benchmarks.

\begin{table}[H]
\centering
\caption{\textbf{Published FireRedLID accuracy (\%).} Results reported by \citet{xu2026fireredasr2s}.}
\label{tab:fireredlid-published}
\footnotesize
\setlength{\tabcolsep}{2pt}
\renewcommand{\arraystretch}{1.02}
\begin{tabular*}{\linewidth}{@{\tableedgepad\extracolsep{\fill}}llrlr@{\tableedgepad}}
\toprule
\tableheaderrow
Benchmark & Coverage & FireRedLID & Baseline & Baseline Acc. \\
\midrule
FLEURS test & 82 langs. & \textbf{97.18} & SpeechBrain & 92.91 \\
CommonVoice test & 74 langs. & \textbf{92.07} & Whisper & 80.81 \\
Chinese dialects & 10+ dialects & \textbf{88.47} & Dolphin & 69.01 \\
\bottomrule
\end{tabular*}
\end{table}

\paragraph{In-house language-routing audit.}
We further compared FireRedLID and SpeechBrain on a nine-language subset of
the internal routing pool. Both systems abstain below a confidence threshold,
so Table~\ref{tab:lid-internal} reports accuracy on retained predictions
together with the number of retained items. At the $\tau=0.8$ operating point
used by the data filter, FireRedLID raises aggregate accuracy from 86.9\% to
93.5\% while retaining more items (649 vs.\ 616), and it is more accurate.

\begin{table}[H]
\centering
\caption{\textbf{In-house nine-language LID audit at confidence $\tau\geq0.8$.}
Accuracy is computed on retained predictions. Parentheses give retained item
counts from the same evaluation pool. Bold marks the higher per-language
accuracy, not greater coverage.}
\label{tab:lid-internal}
\footnotesize
\setlength{\tabcolsep}{5pt}
\renewcommand{\arraystretch}{1.00}
\begin{tabular*}{\linewidth}{@{\tableedgepad\extracolsep{\fill}}lrrr@{\tableedgepad}}
\toprule
\tableheaderrow
Language & SpeechBrain Acc. (kept) & FireRedLID Acc. (kept) & $\Delta$ Acc. \\
\midrule
Bengali (bn)    & 83.3 (54)          & \textbf{92.1} (63)  & $+8.8$ \\
German (de)     & \textbf{89.5} (76) & 87.3 (63)           & $-2.2$ \\
English (en)    & 98.9 (93)          & \textbf{100.0} (95) & $+1.1$ \\
Spanish (es)    & 71.7 (53)          & \textbf{91.4} (70)  & $+19.7$ \\
French (fr)     & 88.9 (81)          & \textbf{90.3} (72)  & $+1.4$ \\
Japanese (ja)   & 87.7 (73)          & \textbf{95.3} (86)  & $+7.6$ \\
Korean (ko)     & 94.0 (83)          & \textbf{95.7} (69)  & $+1.7$ \\
Russian (ru)    & 69.2 (52)          & \textbf{90.2} (51)  & $+21.0$ \\
Chinese (zh)    & 82.4 (51)          & \textbf{95.0} (80)  & $+12.6$ \\
\midrule
Overall         & 86.9 (616)         & \textbf{93.5} (649) & $+6.6$ \\
\bottomrule
\end{tabular*}
\end{table}

\subsection{Transcription}
\label{app:asr-validation}

\paragraph{Published benchmarks.}
Table~\ref{tab:qwen3-asr-published} compares Qwen3-ASR-1.7B with
Whisper-large-v3 (Whisper-L3). FLEURS-en uses WER, FLEURS-zh uses CER, and the
remaining rows use Macro Err., the unweighted mean of language-appropriate
WER or CER.

\begin{table}[H]
\centering
\caption{\textbf{Published Qwen3-ASR-1.7B error rates (\%).} Results reported by \citet{shi2026qwen3}.}
\label{tab:qwen3-asr-published}
\footnotesize
\setlength{\tabcolsep}{2pt}
\renewcommand{\arraystretch}{1.02}
\begin{tabular*}{\linewidth}{@{\tableedgepad\extracolsep{\fill}}llrr@{\tableedgepad}}
\toprule
\tableheaderrow
Benchmark & Coverage & Qwen3-ASR-1.7B & Whisper-L3 \\
\midrule
FLEURS-en & English & \textbf{3.35} & 4.08 \\
FLEURS-zh & Chinese & \textbf{2.41} & 4.09 \\
MLS & 8 langs. & \textbf{8.55} & 8.62 \\
CommonVoice & 13 langs. & \textbf{9.18} & 10.77 \\
FLEURS & 12 langs. & \textbf{4.90} & 5.27 \\
FLEURS extended & 30 langs. & 12.60 & \textbf{8.16} \\
\bottomrule
\end{tabular*}
\end{table}

Qwen3-ASR-1.7B improves on English, Chinese, and multilingual suites.
On the Vaani-V1.0 Hindi leaderboard, IndicConformer-600M achieves 13.2\% WER
versus 26.8\% for Whisper-L3 and is within 0.2 points of the best listed
result (13.0\%) \citep{ai4bharat2025indicconformer,pulikodan2026vaani}.
Vaani-V1.0 contains 5,050 Hindi segments, including code-switched speech,
from 1,103 speakers (11~h).

\paragraph{Deployment check.}
To confirm that our deployment configuration preserves the published accuracy,
we ran the deployed Qwen3-ASR-1.7B streaming implementation (vLLM, 2-s chunks)
on the full LibriSpeech test-clean split (2,620 utterances, 5~h). It obtains
2.35\% WER with an RTF of 0.038 and a mean latency of 283~ms.

\subsection{Speech Segmentation}
\label{app:segmentation-validation}

A streaming-VAD check against offline reference boundaries yields a frame F1
of 0.965, a segment IoU of 0.929, start/end boundary errors of 23.6/24.7~ms,
and an RTF of 0.0077. Speaker assignments and turn boundaries are further
verified manually for all 118 benchmark cases, which span up to 30~min and four
speakers (Appendix~\ref{app:testset}).

\subsection{Expressive Attribute Extractors}
\label{app:attribute-validation}

\paragraph{Gender.}
Table~\ref{tab:gender-selection} compares classifiers on clean FLEURS read
speech in English, Chinese,
French, German, Spanish, Russian, Arabic, Hindi, and Japanese.
We use wav2vec2 (wav2vec2-large-robust-24-ft-age-gender),
which achieved 0.936 accuracy at an RTF of 0.0015 and also provides the
child class used by the age cascade. WavLM (wavlm-large-age-sex) was a
competitive gender candidate (0.938 accuracy), but its gender output is not
used in the annotation pipeline.

\begin{table}[H]
\centering
\caption{\textbf{Gender classification on FLEURS (9 languages).} Bold marks the
deployed classifier. Higher Acc.\ and lower RTF are better. RTF is processing
time divided by audio duration.}
\label{tab:gender-selection}
\small
\setlength{\tabcolsep}{6pt}
\renewcommand{\arraystretch}{1.05}
\vspace{2pt}
\begin{tabular*}{0.70\linewidth}{@{\extracolsep{\fill}}lrr@{}}
\toprule
\tableheaderrow
Method & Acc. & RTF \\
\midrule
WavLM & 0.938 & 0.0028 \\
\textbf{wav2vec2} & \textbf{0.936} & \textbf{0.0015} \\
Qwen2-Audio & 0.919 & 0.0077 \\
\bottomrule
\end{tabular*}
\end{table}

\paragraph{Mandarin accent.}
Table~\ref{tab:accent-selection} compares the deployed
FireRedLID$\rightarrow$Voxlect cascade with its components on the seven-class
KeSpeech split. The cascade achieved 0.757 accuracy and 0.743 UAR, compared
with 0.743/0.733 for Voxlect (Whisper-L3) and 0.331/0.243 for FireRedLID
alone. It improved both overall accuracy and class-balanced recall over
either component.

\begin{table}[H]
\centering
\caption{\textbf{Mandarin-accent classification on KeSpeech (7 classes).} Bold marks
the deployed method. Higher Acc.\ and UAR are better.}
\label{tab:accent-selection}
\small
\setlength{\tabcolsep}{6pt}
\renewcommand{\arraystretch}{1.05}
\vspace{2pt}
\begin{tabular*}{0.70\linewidth}{@{\extracolsep{\fill}}lrr@{}}
\toprule
\tableheaderrow
Method & Acc. & UAR \\
\midrule
\textbf{FireRedLID $\rightarrow$ Voxlect} & \textbf{0.757} & \textbf{0.743} \\
Voxlect (Whisper-L3) & 0.743 & 0.733 \\
FireRedLID & 0.331 & 0.243 \\
\bottomrule
\end{tabular*}
\end{table}

\paragraph{Speaking rate.}
Table~\ref{tab:speaking-rate-selection} compares transcript character, phone,
and word counts divided by each clip's total duration. Five-fold
cross-validation on 246 ParaSpeechCaps (PSC) \citep{diwan2025scaling} clips used three labels: slow (60), measured
(168), and fast (18). Characters/s achieved 0.894 accuracy and 0.937 UAR,
compared with 0.732/0.830 for phones/s and 0.732/0.819 for words/s. The
higher UAR matters on this unevenly distributed testset, where most clips
are labeled measured. We therefore use characters per second for the
deployed speaking-rate annotation, interpreting the value according to
language and transcript format.

\begin{table}[H]
\centering
\caption{\textbf{Speaking-rate measures on PSC ($n=246$).} Bold marks the deployed
measure. Higher Acc.\ and UAR are better.}
\label{tab:speaking-rate-selection}
\small
\setlength{\tabcolsep}{6pt}
\renewcommand{\arraystretch}{1.05}
\vspace{2pt}
\begin{tabular*}{0.70\linewidth}{@{\extracolsep{\fill}}lrr@{}}
\toprule
\tableheaderrow
Measure & Acc. & UAR \\
\midrule
\textbf{Characters/s} & \textbf{0.894} & \textbf{0.937} \\
Phones/s & 0.732 & 0.830 \\
Words/s & 0.732 & 0.819 \\
\bottomrule
\end{tabular*}
\end{table}

\paragraph{Age.}
Age was examined on an internal testset of 185 English TTS clips and 200
Chinese clips, all with human-verified age labels.
We compare wav2vec2
(wav2vec2-large-robust-24-ft-age-gender), WavLM (wavlm-large-age-sex), their
median, and Qwen3-Omni-Thinking over five age bands ($<13$, 13--17, 18--42,
43--59, and $\geq60$). The standalone regressors and their median are
diagnostics. The deployed age tag additionally uses the wav2vec2 gender head's
child class. If that head predicts child, the tag is child regardless of the
regressed ages. Otherwise, we bin the median of the two continuous estimates
using boundaries at 13, 18, 43, and 60 years. For long recordings, each
estimate is first averaged across segments with duration weights, then the
same rule assigns one age band. The captioner receives this perceived age
band, while both estimates remain in the annotation record.

\paragraph{Emotion.}
We calibrate the confidence gate of the in-house seven-class emotion
recognizer on four natural-speech sets (Table~\ref{tab:emotion-gating}).
Across these four sets, the 0.7 gate raises WF1 from 0.45--0.56 to
0.66--0.77 while retaining 20--29\% of examples. A 0.9 gate retains only
3--7\%, so we use 0.7 and leave lower-confidence emotion intervals unlabeled
rather than forcing a class.

\begin{table}[H]
\centering
\caption{\textbf{Seven-class emotion recognition with and without confidence gating.}
Gated scores are computed on retained clips.}
\label{tab:emotion-gating}
\small
\setlength{\tabcolsep}{3pt}
\renewcommand{\arraystretch}{1.08}
\begin{tabular*}{\linewidth}{@{\tableedgepad\extracolsep{\fill}}lc*{7}{r}@{\tableedgepad}}
\toprule
\tableheaderrow
\multirow{2}{*}{Set} & \multirow{2}{*}{Lang.} & \multirow{2}{*}{$n$}
& \multicolumn{3}{c}{No gate} & \multicolumn{3}{c@{\tableedgepad}}{$\tau=0.7$} \\
\cmidrule(lr){4-6}\cmidrule(l){7-9}
\tablesubheaderrow
& & & WA & UA & WF1 & WA & UA & WF1 \\
\midrule
MSP-Podcast & en & 3,256 & 0.49 & 0.45 & 0.45 & 0.72 & 0.54 & 0.66 \\
MELD & en & 2,608 & 0.53 & 0.39 & 0.53 & 0.71 & 0.52 & 0.71 \\
\addlinespace[2pt]
M3ED & zh & 4,194 & 0.53 & 0.42 & 0.54 & 0.73 & 0.59 & 0.73 \\
BIIC-Podcast & zh & 10,149 & 0.57 & 0.37 & 0.56 & 0.79 & 0.55 & 0.77 \\
\bottomrule
\end{tabular*}
\end{table}

\paragraph{Summary.}
These diagnostics support three design decisions. The wav2vec2 gender head
approaches the strongest tested accuracy at substantially lower RTF and
provides the child class used for age annotation.
The FireRedLID$\rightarrow$Voxlect cascade improves Mandarin accent recognition
over either component alone. The 0.7 emotion threshold trades coverage for
substantially higher retained-set WF1. For age, the annotation pipeline
combines the wav2vec2 child class with the median of the two age regressions
and passes the resulting perceived age band to the captioner.

\section{Dataset Statistics}
\label{app:statistics}

\subsection{Source Composition}
\label{app:source-profile}

\begin{figure}[H]
\centering
\includegraphics[width=0.86\linewidth]{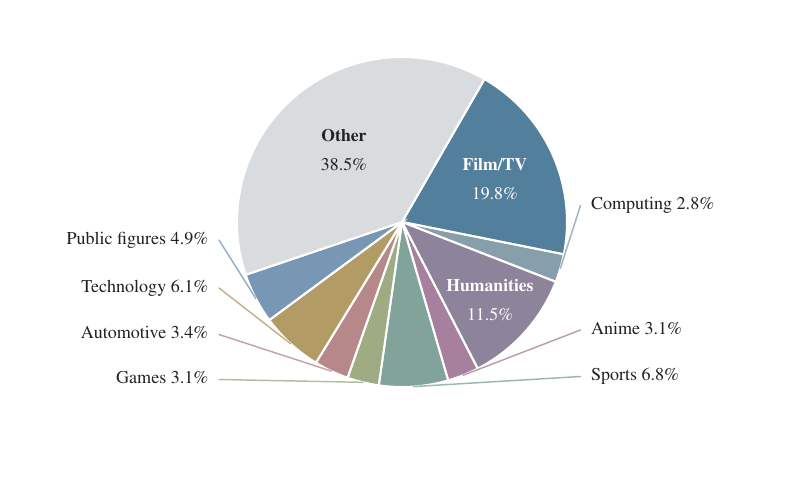}
\caption{\textbf{Topical composition of the source data.} Slice labels report the
share of data hours. Other (38.5\%) combines 25 lower-frequency topics.}
\label{fig:source-domains}
\end{figure}

Figure~\ref{fig:source-domains} shows its topical composition by audio duration:
film/TV commentary accounts for 19.8\%, followed by
humanities (11.5\%), sports (6.8\%), technology (6.1\%), and public figures
(4.9\%). Other accounts for 38.5\%, combining 25
lower-frequency topics.

\Needspace{6\baselineskip}
\subsection{Duration Distribution}

Table~\ref{tab:duration-bins} reports the example-duration distributions of the
single-speaker long-form and dialogue views. Examples longer than five minutes
contribute 50.0\% of \longset hours and 66.8\% of \dialogueset hours, and
examples longer than 30~min alone contribute 11.9\% and 23.7\%, respectively.

\begin{table}[H]
\centering
\caption{\textbf{Example-duration distributions of \longset and \dialogueset.} Each view
is counted separately. Hours and percentages are rounded, so their sums may
differ slightly from the totals.}
\label{tab:duration-bins}
\footnotesize
\setlength{\tabcolsep}{3pt}
\renewcommand{\arraystretch}{0.95}
\begin{tabular*}{\linewidth}{@{\tableedgepad\extracolsep{\fill}}lrrrrrrrr@{\tableedgepad}}
\toprule
\tablesubheaderrow
\multirow{2}{*}{Duration} & \multicolumn{4}{c}{\longset} & \multicolumn{4}{c@{\tableedgepad}}{\dialogueset} \\
\cmidrule(lr){2-5}\cmidrule(l){6-9}
\tablesubheaderrow
& Examples & (\%) & Hours & (\%) & Examples & (\%) & Hours & (\%) \\
\midrule
30--60 s & 2,337,079 & 54.81 & 27,015 & 17.35 & 557,474 & 36.24 & 6,696 & 6.94 \\
1--5 min & 1,571,270 & 36.85 & 50,868 & 32.67 & 737,290 & 47.93 & 25,321 & 26.25 \\
5--10 min & 196,474 & 4.61 & 22,972 & 14.76 & 114,598 & 7.45 & 13,441 & 13.94 \\
10--15 min & 73,138 & 1.72 & 14,827 & 9.52 & 47,164 & 3.07 & 9,589 & 9.94 \\
15--20 min & 33,896 & 0.79 & 9,722 & 6.24 & 25,552 & 1.66 & 7,363 & 7.63 \\
20--25 min & 18,373 & 0.43 & 6,825 & 4.38 & 16,794 & 1.09 & 6,243 & 6.47 \\
25--30 min & 10,977 & 0.26 & 4,998 & 3.21 & 10,908 & 0.71 & 4,975 & 5.16 \\
$>$30 min & 22,706 & 0.53 & 18,461 & 11.86 & 28,459 & 1.85 & 22,824 & 23.66 \\
\midrule
\tabletotalrow Total & 4,263,913 & 100.00 & 155,688 & 100.00 & 1,538,239 & 100.00 & 96,452 & 100.00 \\
\bottomrule
\end{tabular*}
\end{table}

\subsection{Language Distribution}

Table~\ref{tab:language-distribution} shows the seven largest \shortset
languages across all three views. Other groups the remaining languages. Each example is assigned to one language. When
its ordered label contains multiple languages, we use the first, dominant
language by within-example coverage.

\begin{table}[H]
\centering
\caption{\textbf{Language distribution of the three training views.} Each example is
assigned to its dominant language; Other aggregates the remaining languages. Long and Dialogue hours are uniformly
normalized to the corresponding totals in Table~\ref{tab:tracks}. Hours and
shares are rounded.}
\label{tab:language-distribution}
\footnotesize
\setlength{\tabcolsep}{2pt}
\renewcommand{\arraystretch}{0.95}
\begin{tabular*}{\linewidth}{@{\tableedgepad\extracolsep{\fill}}lrrrrrr@{\tableedgepad}}
\toprule
\tableheaderrow
\multirow{2}{*}{Language} & \multicolumn{2}{c}{\shortset}
& \multicolumn{2}{c}{\longset} & \multicolumn{2}{c@{\tableedgepad}}{\dialogueset} \\
\cmidrule(lr){2-3}\cmidrule(lr){4-5}\cmidrule(l){6-7}
\tablesubheaderrow
& Hours & Share (\%) & Hours & Share (\%) & Hours & Share (\%) \\
\midrule
English & 36,808 & 25.16 & 45,311 & 29.10 & 36,238 & 37.57 \\
Chinese & 36,739 & 25.11 & 54,257 & 34.85 & 17,383 & 18.02 \\
German & 24,646 & 16.84 & 19,703 & 12.66 & 21,701 & 22.50 \\
Japanese & 14,940 & 10.21 & 8,491 & 5.45 & 7,710 & 7.99 \\
Arabic & 8,400 & 5.74 & 7,928 & 5.09 & 3,253 & 3.37 \\
Korean & 7,796 & 5.33 & 5,200 & 3.34 & 2,649 & 2.75 \\
Spanish & 4,523 & 3.09 & 4,194 & 2.69 & 2,341 & 2.43 \\
Other & 12,473 & 8.53 & 10,604 & 6.81 & 5,177 & 5.37 \\
\midrule
\tabletotalrow Total & 146,325 & 100.00 & 155,688 & 100.00 & 96,452 & 100.00 \\
\bottomrule
\end{tabular*}
\end{table}

\subsection{Expressive Annotation Distribution}

\begin{figure}[H]
\centering
\includegraphics[width=\textwidth]{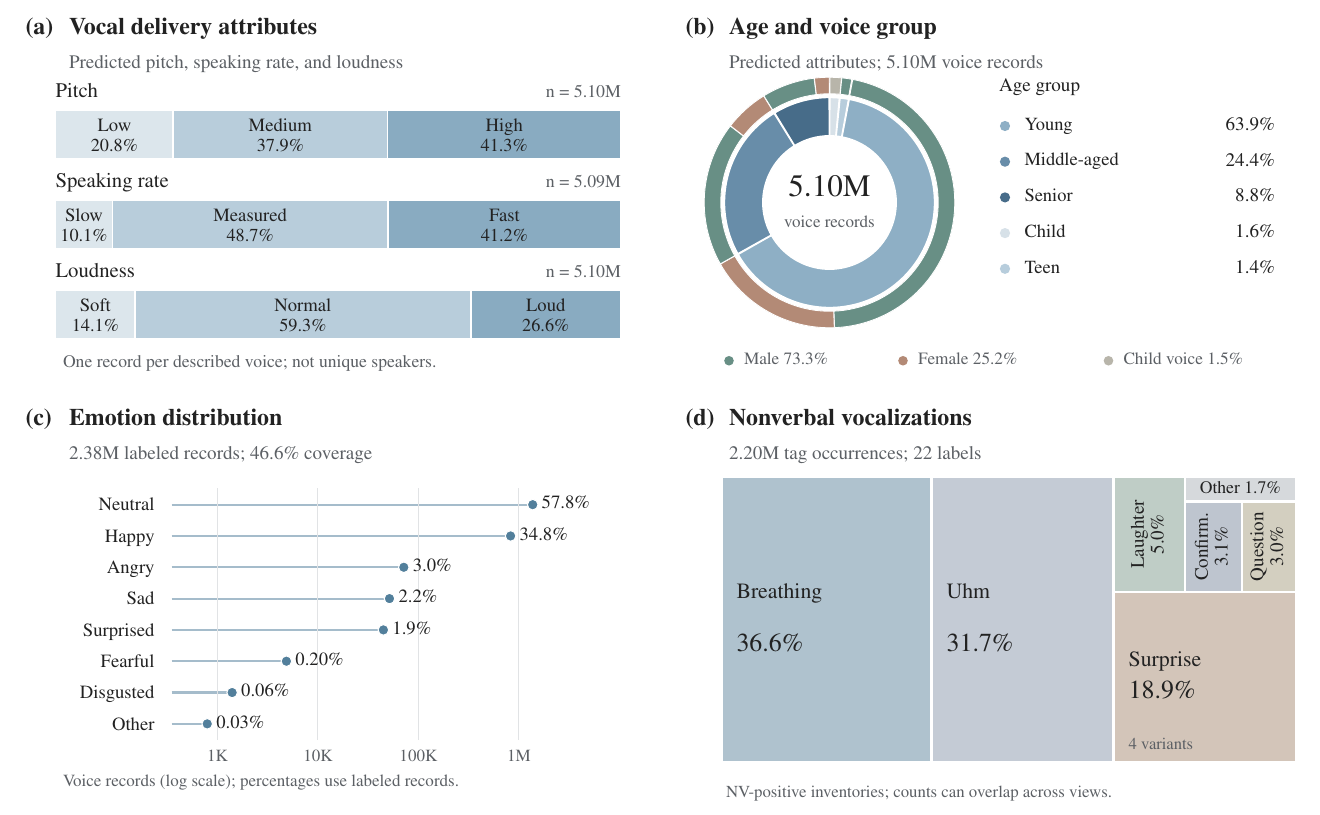}
\caption{\textbf{Expressive annotation distributions of \dataset.}
(a--c) Predicted attributes over 5,096,107 voice records from 4,766,441
instruction-annotated examples, with one record per speaker in Dialogue.
Attribute distributions exclude missing labels.
In (b), age forms the inner ring and voice-group labels the outer ring.
(d) NV tag occurrences in 933,006 NV-positive records, grouping Surprise and
Question variants. Records and occurrences may overlap across views.}
\label{fig:expressive-statistics}
\end{figure}

Figure~\ref{fig:expressive-statistics} summarizes the distributions of
predicted voice attributes and NV tags. Table~\ref{tab:tracks} reports the
corresponding per-view dataset: audio hours count complete examples,
including intervening pauses, and NV hours count the full duration of each
tagged example. Totals sum view entries.
Because each instruction is grounded in task-specific attribute predictions
and their temporal changes, and NV tags come from a dedicated recognizer,
expressive supervision in \dataset is tied to measurable acoustic evidence
rather than relying solely on end-to-end speech-language-model captioning.

\section{\dataset-testset Details}
\label{app:testset}

This appendix complements Section~\ref{subsec:testset} with the construction
and verification procedure, coverage statistics, stratified generation
analysis, a comparison with existing testsets, and the construction of the
long-form control evaluation sets.
Evaluation metrics and listening protocols are specified in
Appendix~\ref{app:eval-protocol}.

\subsection{Construction and Verification}

\paragraph{Construction and speaker references.}
Candidates are drawn from continuous source recordings and pass the
acoustic, language, and diarization stages of \emiliapipe. Ten annotators
annotate and verify transcripts, speaker assignments, turn boundaries, audio
integrity, and the voice identity in each prompt. A second annotator
independently reviews 10\% of the annotations and directly corrects clear
errors. Annotations with substantial errors are reassigned to another
annotator. The target script preserves turn order and explicit speaker
identifiers, linking each turn to its reference voice. Generation is
conditioned on this script and the per-speaker prompt audio and transcripts,
and the natural target recording serves as the evaluation reference. Each of
the 118 cases uses a distinct source recording.

\paragraph{Train--test isolation.}
For \dataset-testset, complete test sources and detected duplicate groups,
including derived views and prompts, are excluded from training. Across all
evaluations, we remove Internet-scale training sources whose URLs match
available target- or prompt-audio source URLs. TTSD-Eval
\citep{zhang2026moss} uses scripts and speaker prompts; Long-TTS-Eval
\citep{wang2026mgm} uses target text and Seed-TTS-Eval prompts. Neither uses
target ground-truth audio as an inference reference. Seed-TTS-Eval
\citep{anastassiou2024seed} uses Common Voice (English) and DiDiSpeech-2
(Chinese); MiniMax-Multilingual-24 \citep{zhang2025minimax} uses Common Voice
speaker prompts. URL screening establishes source-URL nonoverlap, while
duplicate-group exclusion also covers detected copies of our test sources.

\paragraph{Audio records and consistency checks.}
Case selection is stratified by language, duration, and speaker count. The
dataset includes 118 target recordings and 241 speaker prompts, all stored as
mono 44.1~kHz FLAC. Scripts contain 3--225 turns, and speaker prompts span
3.55--29.98~s (16.44~s on average).

\subsection{Coverage and Comparison with Existing testsets}
\label{app:testset-stratified}

\paragraph{Duration and speaker coverage.}
As summarized in Table~\ref{tab:testset-composition} and Table~\ref{tab:testset-stratified-counts}, the testset contains
42 single-speaker long-form cases and 76 dialogue cases, evenly split between
Chinese and English (59 each) and totaling 23~h. Target durations range from
51.1~s to 30.0~min and average 11.85~min. For the stratified analysis,
32, 22, 21, 18, 15, and 10 cases are assigned to the $<$ 5, 5--10, 10--15,
15--20, 20--25, and 25--30~min groups, respectively. The numbers of cases with one, two, three, and four
speakers are 42, 40, 25, and 11.

\paragraph{Comparison.}
Table~\ref{tab:long-dialogue-benchmarks} compares \dataset-testset with
existing single-speaker long-form and dialogue evaluation resources. It is the
only open resource that provides per-speaker prompt audio for recordings
averaging over 10~min with up to four speakers.

\begin{table}[tb]
    \centering
    \caption{\textbf{Comparison of long-form TTS testsets.} Average duration is per
    target recording, using published or ground-truth audio durations when available.
    $^\dagger$ denotes an estimate from target text at 3 characters per
    second.  ``Prompt audio'' denotes a reference recording for each speaker.}
    \label{tab:long-dialogue-benchmarks}
    \small
    \setlength{\tabcolsep}{2.5pt}
    \renewcommand{\arraystretch}{0.95}

    \begin{tabular*}{\linewidth}{@{\tableedgepad\extracolsep{\fill}}P{0.42\linewidth}crrccc@{\tableedgepad}}
    \toprule
    \tableheaderrow
    Benchmark & Type & Clips & \shortstack{Avg.\\dur.} & Speaker & \shortstack{Open\\source} & \shortstack{Prompt\\audio} \\
    \midrule
    MinutesSpeech-test \citep{nishimura2025hall}
        & Long & 1,221 & 0.84 min & 1 & \no & \yes \\
    LibriSpeech-Long \citep{park2024long}
        & Long & 960 & 3.19 min & 1 & \yes & \no \\
    Long-TTS-Eval \citep{wang2026mgm}
        & Long & 694 & 13.4 min$^\dagger$ & 1 & \yes & \no \\
    MultiDialog \citep{park2024let}
        & Dialogue & 831 & 2.46 min & 2 & \yes & \no \\
    ZipVoice-Dialog testset \citep{zhu2026zipvoice}
        & Dialogue & 637 & 0.38 min & 2 & \yes & \yes \\
    TTSD-Eval \citep{zhang2026moss}
        & Both & 100 & 10.0 min$^\dagger$ & 2 & \yes & \yes \\
    SwanBench-Speech \citep{pan2026comprehensive}
        & Both & 1,101 & 3.0 min$^\dagger$ & 1--4 & \no & \no \\
    VIBEVOICE-Eval \citep{peng2026vibevoice}
        & Both & 108 & 16.03 min & 1--4 & \no & \yes \\
    \midrule
    \tableoursrow \dataset-testset
        & Both & 118 & 11.85 min & 1--4 & \yes & \yes \\
    \bottomrule
    \end{tabular*}
\end{table}

\begin{table}[H]
\centering
\caption{\textbf{\dataset-testset cases by duration and speaker count.}
Each language contains 59 recordings.}
\label{tab:testset-stratified-counts}
\small
\setlength{\tabcolsep}{4pt}
\begin{tabular}{lrrrrrrrrr}
\toprule
& \multicolumn{4}{c}{English speakers} & \multicolumn{4}{c}{Chinese speakers} & \\
\cmidrule(lr){2-5}\cmidrule(lr){6-9}
Target duration & 1 & 2 & 3 & 4 & 1 & 2 & 3 & 4 & Total \\
\midrule
$<$5 min    & 8 & 4 & 2 & 1 & 5 & 6 & 4 & 2 & 32 \\
5--10 min   & 4 & 2 & 4 & 2 & 3 & 4 & 3 & 0 & 22 \\
10--15 min  & 3 & 4 & 4 & 1 & 2 & 4 & 2 & 1 & 21 \\
15--20 min  & 3 & 4 & 1 & 0 & 3 & 2 & 2 & 3 & 18 \\
20--25 min  & 3 & 2 & 1 & 0 & 3 & 4 & 2 & 0 & 15 \\
25--30 min  & 3 & 3 & 0 & 0 & 2 & 1 & 0 & 1 & 10 \\
\midrule
Total       & 24 & 19 & 12 & 4 & 18 & 21 & 13 & 7 & 118 \\
\bottomrule
\end{tabular}
\end{table}

\begin{figure}[H]
\centering
\includegraphics[width=0.9\linewidth]{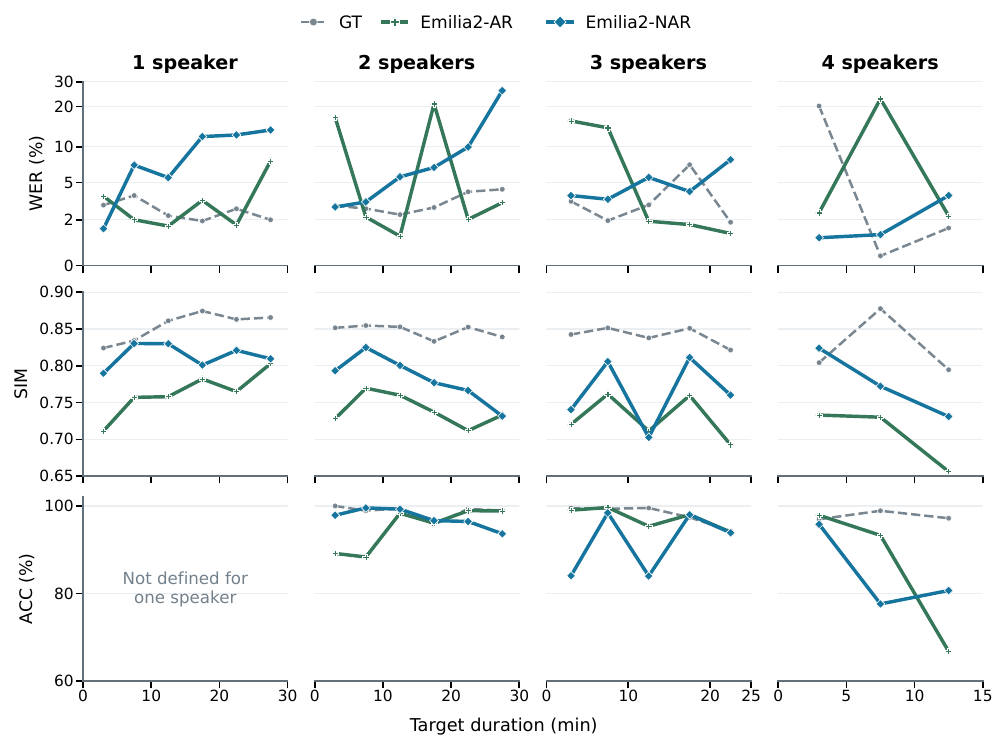}
\caption{\textbf{English WER, prompt SIM, and speaker ACC versus target duration.}
Columns show one through four speakers. Curves compare matched GT,
Emilia2-AR, and Emilia2-NAR recordings. Single-speaker ACC is
undefined. Cell counts are in Table~\ref{tab:testset-stratified-counts}.}
\label{fig:testset-duration-english}
\end{figure}

\begin{figure}[H]
\centering
\includegraphics[width=0.9\linewidth]{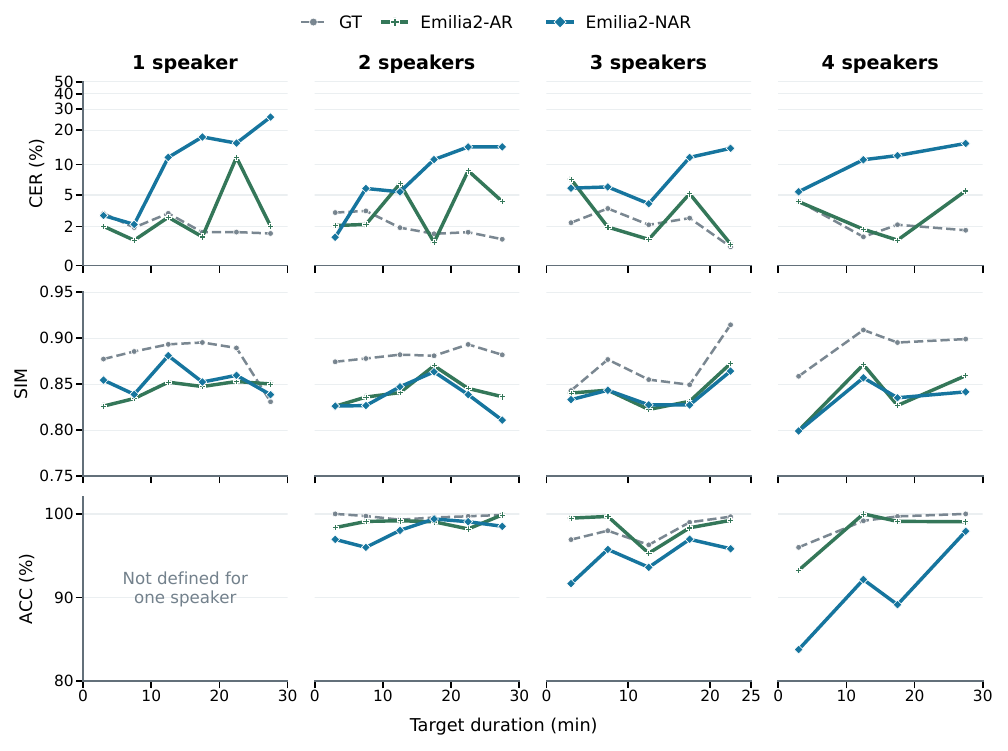}
\caption{\textbf{Chinese CER, prompt SIM, and speaker ACC versus target duration.}
Systems and speaker-count groups follow
Figure~\ref{fig:testset-duration-english}. In the four-speaker column,
markers occur only at
scored bins. Cell counts are in Table~\ref{tab:testset-stratified-counts}.}
\label{fig:testset-duration-chinese}
\end{figure}

Figures~\ref{fig:testset-duration-english} and~\ref{fig:testset-duration-chinese}
compare matched English and Chinese cases by duration and speaker count, with
cell sizes in Table~\ref{tab:testset-stratified-counts}. Emilia2-NAR has larger
content errors in longer recordings, while Emilia2-AR stays closer to GT in
the long two-speaker bins. In paired
single-speaker one-minute windows, prompt SIM at 9--10~min changes by $+0.004$
for GT, and $-0.017$ for NAR relative to the first
minute.

\subsection{Long-Form Control Evaluation Sets}
\label{app:testset-control}

The instruction and NV studies each use a separate evaluation set drawn from
\dataset-testset cases whose target recordings are shorter than 5~min. For the instruction study, control prompts are taken from the \emiliapipe
instruction annotations before evaluation. For the NV
study, DeepSeek rewrites each case script and inserts a number of NV tags
scaled to the target duration.

\section{Model and Training Details}
\label{app:training}

\subsection{Model Objectives}
\label{app:model-objectives}

\begin{figure}[tb]
\centering
\includegraphics[width=\textwidth]{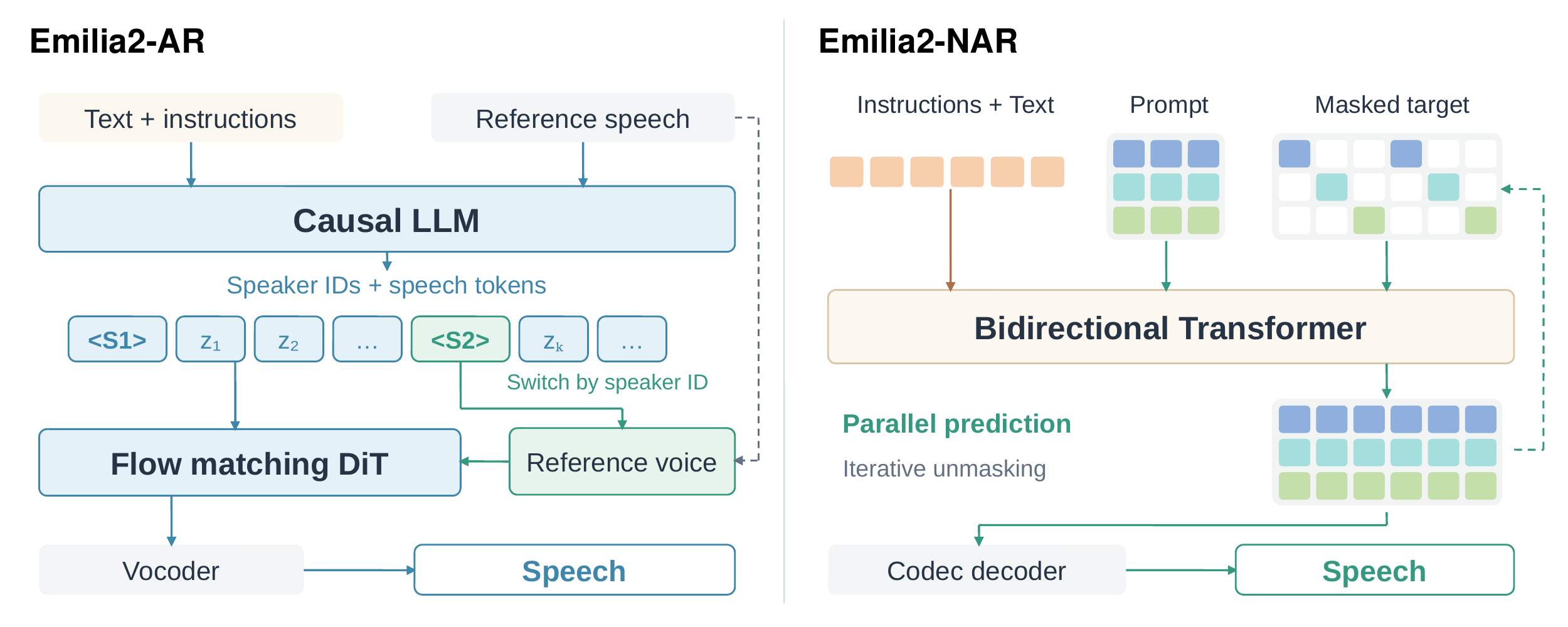}
\caption{\textbf{Emilia2 model architectures.} (a) Emilia2-AR combines autoregressive token prediction with
flow-matching reconstruction. Generated speaker IDs select the associated
reference voice. (b) Emilia2-NAR learns masked acoustic-token prediction
and generates speech through iterative parallel unmasking.}
\label{fig:model-architectures}
\end{figure}

\paragraph{Emilia2-AR.}
Equation~\ref{eq:arfm-factorization} is optimized by minimizing the negative
log-likelihood of the observed sequence.
Reference speech and its transcript provide an in-context voice example.
Ordered dialogue turns supervise speaker-ID transitions, while instructions
and NV annotations specify style conditions and nonverbal content. A
conditional flow-matching decoder generates a mel spectrogram from speech
tokens. A Pupu-Vocoder \citep{gu2026aliasing} renders it as a waveform.
During dialogue generation, each emitted speaker-ID marker selects the
corresponding reference voice for subsequent reconstruction.

\paragraph{Emilia2-NAR.}
The token-level objective in Equation~\ref{eq:nar-token-objective} is aggregated
over masked entries with codebook-specific weights. Visible target tokens
provide bidirectional acoustic context, while the script and reference voices
supply content and speaker conditioning.

We sample one masking strategy per example: full masking with probability
30\%, coarse-to-fine masking with probability 50\%, and independent per-cell
masking with probability 20\%. The partially masked strategies draw an
example-level ratio $r\sim\mathcal{U}(0,1)$. Full masking sets $M_{k,t}=1$
for all target entries, whereas per-cell masking independently draws
$M_{k,t}\sim\mathrm{Bernoulli}(r)$. For coarse-to-fine masking, we draw
$d\sim\mathcal{U}\{0,\ldots,K-1\}$ and
$B_t\sim\mathrm{Bernoulli}(r)$ and apply
Equation~\ref{eq:nar-depth-mask}. Thus lower-index codebooks remain visible,
the boundary codebook is partially masked, and higher-index codebooks are
hidden, training completion at different depths.

Inference begins with a fully masked target conditioned on the script and
reference speech. At each iteration, the model predicts candidate tokens in
parallel and reveals a confidence-selected subset using a codebook-depth
penalty. Lower-index codebooks receive higher priority, while reference and
previously filled entries remain fixed. An acoustic decoder reconstructs the
waveform after iterative unmasking completes.

\subsection{Training Configuration}

Both models are trained on eight GPUs with BF16 mixed precision. Emilia2-AR
uses AdamW with weight decay 0.01 and gradient-norm clipping at 1.0. Its first
three stages use 3\% warmup followed by cosine decay. Expressive SFT uses a
constant learning rate. AR gradient accumulation is one in the first three
stages and four in expressive SFT. Table~\ref{tab:training-config} gives its
stage-wise learning rates and dynamic per-GPU batch limits. Emilia2-NAR uses
learning rates of $10^{-4}$ in the first three stages and $10^{-5}$ in
expressive SFT. 

\begin{table}[H]
\centering
\caption{\textbf{Representative Emilia2-AR training settings.}}
\label{tab:training-config}
\normalsize
\setlength{\tabcolsep}{5pt}
\renewcommand{\arraystretch}{1.15}
\begin{tabular*}{0.88\linewidth}{@{}l@{\extracolsep{\fill}}ccc@{}}
\toprule
\tableheaderrow
Stage & Max. duration & Learning rate & Max.\ token length \\
\midrule
Short-form & 30~s & $2\times10^{-4}$ & 32,768 \\
Long-form & 30~min & $2\times10^{-4}$ & 32,768 \\
Dialogue & 30~min & $2\times10^{-4}$ & 50,480 \\
Expressive SFT & 5~min & $1\times10^{-5}$ & 50,480 \\
\bottomrule
\end{tabular*}
\end{table}

\section{Evaluation Protocols}
\label{app:eval-protocol}

\subsection{Inference Settings}

\paragraph{Emilia2-AR.}
The causal token predictor receives the target script, any style instruction,
and reference speech with its transcript for each speaker in the script. It
generates speech tokens and speaker-ID markers autoregressively over the
complete script. Each emitted speaker-ID marker selects the corresponding
reference voice for conditional DiT flow-matching reconstruction of the mel
spectrogram; Pupu-Vocoder then renders the waveform. Standard long-form
evaluations use whole-script generation. The separately evaluated chunked
variant is described in Appendix~\ref{app:chunk-vs-whole}.

\paragraph{Emilia2-NAR.}
Generation uses 256 iterations, classifier-free guidance scale 2.0,
time shift 0.1, layer penalty 5, position temperature 5, class temperature 0,
and zero coarse-stage steps. Model weights are stored in FP32 and evaluated
with BF16 autocast. Only speakers present in the target script provide
reference prompts. Target duration is estimated from target and reference
text together with reference codec frames. Ground-truth duration and target
tokens are not used for standard inference. Generated audio is stored as mono 44.1~kHz PCM16.
Each case uses a deterministic seed derived from a base seed of 42 and its
case identifier.

\subsection{Objective Metrics}

\paragraph{TTSD-Eval.}
We follow the official two-speaker dialogue protocol \citep{zhang2026moss}.
Turn boundaries come from the speaker-tagged script rather than diarization:
the spoken text is aligned to the complete waveform with the MMS-FA CTC forced
aligner, and each fragment, delimited by a speaker tag or sentence punctuation,
inherits the most recent speaker tag. Fragments are embedded with the frozen
WeSpeaker SimAM-ResNet100 model and compared with both speaker prompts. SIM is
the cosine similarity to the scripted speaker, and ACC is the fraction of
fragments whose most similar prompt matches the script. Content accuracy is
measured by Whisper-large-v3 on the complete recording after removing bracketed
fields and punctuation. All metrics are macro-averaged over cases within each
language.

\paragraph{Long-TTS-Eval.}
We use the standard long-form splits with the official content-error protocol
\citep{wang2026mgm}. For each case, we randomly sample prompt audio from the
Seed-TTS-Eval prompt-audio pool \citep{anastassiou2024seed}, following the official protocol. Generated audio is split into contiguous 28-s chunks,
transcribed by Whisper-large-v3 (English) or Paraformer-zh (Chinese), and the
concatenated hypothesis is normalized by lowercasing, punctuation removal, and
numeral expansion. Each case keeps the lower error against the released written
and spoken-form references, and edit counts are pooled within each language.
As the official protocol defines only content error, we additionally report SIM
between the complete output and its prompt using the Seed-TTS-Eval WavLM-large
speaker-verification model, macro-averaged over cases. ACC is undefined for
these single-speaker cases.

\paragraph{\dataset-testset automatic evaluation.}
We adapt the TTSD-Eval protocol to recordings of up to 30~min with one to four
speakers. Each recording is converted to 16-kHz mono and split at silence
boundaries detected by FireRedVAD \citep{xu2026fireredasr2s} into contiguous
chunks targeting 25~s and capped at 30~s, without discarding samples. Chunks
are transcribed independently by Qwen3-ASR-1.7B \citep{shi2026qwen3}, and
the concatenated hypothesis gives complete-recording WER or CER after
lowercasing and removing punctuation and bracketed event labels.

Speaker turns are recovered without reference timestamps. A monotonic
edit-distance alignment maps each chunk hypothesis to a contiguous slice of the
speaker-attributed script; slices with at least 50\% exact ASR anchors are
forced-aligned to the chunk, and aligned tokens inherit sentence, turn, and
speaker labels from the script. A segment ends at a speaker or sentence change,
token discontinuity, gap longer than 1~s, or the 30-s limit. Segments shorter
than 0.5~s are excluded from speaker scoring.

For every system and the ground-truth reference, a single frozen
speaker-verification model embeds each segment and the case-specific
prompts from mean-normalized 80-bin filterbanks, with inputs capped at 30~s and
segments shorter than 2~s repeated. For segment embedding $e_{ij}$ of case
$i$, prompt embedding $p_{ik}$ of speaker $k$, and scripted speaker $y_{ij}$,
let $s_{ijk}=(\cos(e_{ij},p_{ik})+1)/2$. Then
\begin{equation}
\mathrm{SIM}_i=\frac{1}{|J_i|}\sum_{j\in J_i}s_{ijy_{ij}},
\qquad
\mathrm{ACC}_i=\frac{1}{|J_i|}\sum_{j\in J_i}
\mathbf{1}\bigl[\operatorname*{arg\,max}_k s_{ijk}=y_{ij}\bigr].
\label{eq:testset-speaker-metrics}
\end{equation}
SIM is macro-averaged over cases, and ACC
(reported in \%) only over multi-speaker cases, since the single-prompt
decision is trivial. All cases are evaluated, with
metrics macro-averaged within each language.

\subsection{Subjective Listening Tests}
\label{app:listening}

The listening tests use the long-form control evaluation sets described in
Appendix~\ref{app:testset-control}. For the instruction study (Table~\ref{tab:instruction-control-results}),
CER/WER measures content accuracy, while listeners rate naturalness (N-MOS) and
instruction adherence (Instr-MOS) on a 1--5 scale. For the NV study
(Table~\ref{tab:nv-control-results}), listeners rate N-MOS and NV Quality,
which covers event type, placement, and contextual style. Scores are reported
separately for Chinese and English, using the same recordings for all measures
within each task--language stratum.

\paragraph{Evaluators and assignment.}
Ten raters participate. Those assigned to English cases have an IELTS score
of at least 6.5. For each target, two raters independently score both
anonymized system outputs with absolute ratings, so systems are paired by
target and rater. Raters hear the complete recording together with its
instruction (instruction cases) or its script and requested events (NV cases).

\paragraph{Scoring criteria.}
All judgments are case-level. \textbf{N-MOS (1--5)} evaluates pronunciation,
prosody, continuity, and the
absence of audible artifacts, independently of control correctness: 5 is fully
natural; 4 is mostly natural with only minor defects; 3 is fair with clearly
noticeable but non-disruptive defects; 2 is mostly unnatural with frequent or
severe defects; and 1 is completely unnatural or largely unintelligible.

\textbf{Instr-MOS (1--5)} grades instruction adherence:
5 fully and consistently realizes every requirement; 4 meets the main
requirements with minor or localized deviations; 3 is partial or inconsistent;
2 is weak and mostly incorrect; and 1 is absent or contradicts the instruction.

\textbf{NV Quality (1--5)} jointly evaluates the requested event type, its
placement relative to the words, and its contextual style: 5 is correct and
convincing with no salient unintended events; 4 has minor timing, intensity, or
realization deviations; 3 is recognizable but contains noticeable errors; 2 has
major omissions, spurious events, or inappropriate placements; and 1 is absent,
unrecognizable, or predominantly incorrect.

\section{Additional Results}
\label{app:supplementary-evaluations}

We complement Section~\ref{sec:experiments} with an analysis of chunked versus
whole-sequence long-form inference, followed by short-form and multilingual
synthesis on Seed-TTS-Eval \citep{anastassiou2024seed}, which covers English
and Chinese, and MiniMax-Multilingual-24 \citep{zhang2025minimax}, which
covers 24 languages. Seed-TTS-Eval reports WER for English, CER for Chinese,
and speaker similarity (SIM-o). MiniMax-Multilingual-24 reports WER and SIM-o
(protocols in Appendix~\ref{app:eval-protocol}). On both benchmarks, we compare
against Higgs Audio v2, VibeVoice-1.5B, and MOSS-TTSD.

\subsection{Chunked versus Whole-Sequence Long-Form Inference}
\label{app:chunk-vs-whole}

We test whether generating a long recording in short chunks affects its
naturalness relative to generating the complete script in one pass. For
Emilia2-AR, chunked inference splits the tagged script at punctuation and
speaker-turn boundaries into chunks of at most 120 Chinese or 350 English
characters (about 25~s), samples each chunk independently, and concatenates
the codec tokens before waveform decoding. Emilia2-NAR uses 256 sampling
steps in both modes. Ten annotators each rate five Chinese and five English
randomly selected \dataset-testset cases. For each case, the four generated
versions and ground truth are anonymized, randomized, and scored with
case-level N-MOS (1--5 in 0.5-point steps; Appendix~\ref{app:listening}).

\begin{table}[tb]
\centering
\caption{\textbf{Chunked versus whole-sequence inference on \dataset-testset.} N-MOS is reported as the case mean $\pm$ 95\%
confidence interval. Bold marks the better result within each model and
language.}
\label{tab:chunk-vs-whole-results}
\small
\setlength{\tabcolsep}{3pt}
\renewcommand{\arraystretch}{0.95}
\begin{tabular*}{\linewidth}{@{\tableedgepad\extracolsep{\fill}}llrrrrr@{\tableedgepad}}
\toprule
\tableheaderrow
\multirow{2}{*}{Model} & \multirow{2}{*}{Inference}
& \multicolumn{2}{c}{Chinese} & \multicolumn{2}{c}{English}
& \multirow{2}{*}{\shortstack{Overall\\N-MOS$\uparrow$}} \\
\cmidrule(lr){3-4}\cmidrule(lr){5-6}
\tablesubheaderrow
& & CER$\downarrow$ & N-MOS$\uparrow$ & WER$\downarrow$ & N-MOS$\uparrow$ & \\
\midrule
\tabletotalrow Ground truth & Recorded & -- & 4.55 $\pm$ 0.31 & -- & 4.55 $\pm$ 0.27 & 4.55 $\pm$ 0.20 \\
\midrule
\multirow{2}{*}{Emilia2-AR}
& Chunked & 4.36 & 3.30 $\pm$ 0.16 & \textbf{2.44} & 4.08 $\pm$ 0.28 & 3.69 $\pm$ 0.24 \\
& Whole sequence & \textbf{3.67} & \textbf{4.05 $\pm$ 0.23} & 6.82 & \textbf{4.10 $\pm$ 0.24} & \textbf{4.08 $\pm$ 0.16} \\
\midrule
\multirow{2}{*}{Emilia2-NAR}
& Chunked & \textbf{1.73} & 3.90 $\pm$ 0.28 & \textbf{1.52} & 4.00 $\pm$ 0.21 & 3.95 $\pm$ 0.17 \\
& Whole sequence & 8.62 & \textbf{4.05 $\pm$ 0.31} & 6.62 & \textbf{4.10 $\pm$ 0.38} & \textbf{4.08 $\pm$ 0.24} \\
\bottomrule
\end{tabular*}
\end{table}

Whole-sequence generation yields higher full-recording naturalness than stitching independently generated chunks. Chunking lowers CER/WER in three of the four model--language pairs, most markedly for Emilia2-NAR in Chinese (8.62\% to 1.73\%), but lowers N-MOS in all four, with the largest drop observed for Emilia2-AR in Chinese (4.05 to 3.30). These opposite trends show that chunking improves local content accuracy at the expense of long-range continuity: generating each chunk independently resets prosody and turn-taking rhythm at its boundaries. Consequently, CER/WER, SIM, and ACC alone cannot distinguish coherent whole-sequence synthesis from locally accurate but perceptually discontinuous stitching; long-form evaluation must also assess naturalness over the complete recording. \dataset-testset provides paired ground-truth recordings averaging 11.85 min and extending to 30 min, enabling the development and calibration of metrics for continuity over substantially longer horizons.

\subsection{Short-Form Synthesis}

Table~\ref{tab:seed-results} summarizes the results on Seed-TTS-Eval. Emilia2-NAR achieves the lowest error rates of all systems, with an English WER of 1.60\% and a Chinese CER of 1.00\%, both below those of the ground-truth recordings (2.14\% and 1.25\%). Emilia2-AR ranks second in English WER (1.85\%). Emilia2-NAR also outperforms Emilia2-AR in speaker similarity (English/Chinese SIM-o of 0.66/0.75 vs.\ 0.61/0.69), with Chinese SIM-o within 0.01 of the best system.

\begin{table}[H]
\centering
\caption{\textbf{Short-form results on Seed-TTS-Eval.} Bold
marks the best available generated-speech score in each column.}
\label{tab:seed-results}
\small
\setlength{\tabcolsep}{3pt}
\renewcommand{\arraystretch}{0.95}

\begin{tabular*}{\linewidth}{@{\tableedgepad\extracolsep{\fill}}lrrrr@{\tableedgepad}}
\toprule
\tableheaderrow
\multirow{2}{*}{Model} & \multicolumn{2}{c}{English} & \multicolumn{2}{c@{\tableedgepad}}{Chinese} \\
\cmidrule(lr){2-3}\cmidrule(l){4-5}
\tablesubheaderrow
& SIM-o $\uparrow$ & WER $\downarrow$ & SIM-o $\uparrow$ & CER $\downarrow$ \\
\midrule
\tabletotalrow Ground-truth & 0.73 & 2.14 & 0.76 & 1.25 \\
\midrule
Higgs Audio v2 & 0.65 & 2.60 & 0.72 & 2.62 \\
VibeVoice-1.5B & \textbf{0.69} & 3.04 & 0.74 & 1.16 \\
MOSS-TTSD & 0.68 & 9.83 & \textbf{0.76} & 5.84 \\
\midrule
\tableoursrow \textbf{Emilia2-AR} & 0.61 & 1.85 & 0.69 & 1.46 \\
\tableoursrow \textbf{Emilia2-NAR} & 0.66 & \textbf{1.60} & 0.75 & \textbf{1.00} \\
\bottomrule
\end{tabular*}

\end{table}

\subsection{Multilingual Synthesis}

Table~\ref{tab:multilingual-baselines} summarizes the comparison on
MiniMax-Multilingual-24, with per-language Emilia2 results in
Table~\ref{tab:multilingual-results}. Both Emilia2 models reduce the
macro-average WER by more than 29 points relative to every baseline (7.67\% and
10.34\% vs.\ 40.25--85.41\%), reflecting the broad language coverage of
\dataset. Emilia2-NAR's macro-average SIM-o (0.79) is within 0.01 of the best
baseline and exceeds that of Emilia2-AR (0.71) in all 24 languages, whereas
Emilia2-AR attains lower WER in 17 languages.

\begin{table}[H]
\centering
\caption{\textbf{Macro-averaged results on MiniMax-Multilingual-24.} Languages are
weighted equally. WER is in percent. Bold marks the best score per
metric.}
\label{tab:multilingual-baselines}
\small
\renewcommand{\arraystretch}{0.95}
\begin{tabular*}{\linewidth}{@{\tableedgepad\extracolsep{\fill}}lrr@{\tableedgepad}}
\toprule
\tableheaderrow
Model & WER $\downarrow$ & SIM-o $\uparrow$ \\
\midrule
Higgs Audio v2 & 49.51 & 0.70 \\
VibeVoice-1.5B & 85.41 & 0.75 \\
MOSS-TTSD & 40.25 & \textbf{0.80} \\
\midrule
\tableoursrow \textbf{Emilia2-AR} & \textbf{7.67} & 0.71 \\
\tableoursrow \textbf{Emilia2-NAR} & 10.34 & 0.79 \\
\bottomrule
\end{tabular*}
\end{table}

\begin{table}[H]
\centering
\caption{\textbf{Per-language Emilia2 results on MiniMax-Multilingual-24.} WER is in percent recomputed from the 24 displayed language rows. Bold marks the best value for each metric in each row.}
\label{tab:multilingual-results}
\small
\setlength{\tabcolsep}{3pt}
\renewcommand{\arraystretch}{0.95}

\begin{tabular*}{\linewidth}{@{\tableedgepad\extracolsep{\fill}}lrrrr@{\tableedgepad}}
\toprule
\tableheaderrow
\multirow{2}{*}{Language} & \multicolumn{2}{c}{Emilia2-AR} & \multicolumn{2}{c}{Emilia2-NAR} \\
\cmidrule(lr){2-3}\cmidrule(l){4-5}
\tablesubheaderrow
& WER $\downarrow$ & SIM-o $\uparrow$ & WER $\downarrow$ & SIM-o $\uparrow$ \\
\midrule
Arabic     & \textbf{7.34}  & 0.71 & 13.79 & \textbf{0.76} \\
Cantonese  & 37.15          & 0.75 & \textbf{28.21} & \textbf{0.78} \\
Chinese    & 3.72           & 0.73 & \textbf{1.25}  & \textbf{0.79} \\
Czech      & \textbf{11.97} & 0.70 & 18.90 & \textbf{0.76} \\
Dutch      & \textbf{4.47}  & 0.69 & 6.91  & \textbf{0.72} \\
English    & \textbf{1.26}  & 0.73 & 3.23  & \textbf{0.83} \\
Finnish    & \textbf{15.27} & 0.78 & 30.59 & \textbf{0.79} \\
French     & \textbf{3.52}  & 0.62 & 4.69  & \textbf{0.73} \\
German     & \textbf{0.55}  & 0.71 & 2.96  & \textbf{0.77} \\
Greek      & 8.52           & 0.75 & \textbf{7.00}  & \textbf{0.83} \\
Hindi      & 20.96          & 0.72 & \textbf{6.17}  & \textbf{0.84} \\
Indonesian & \textbf{12.30} & 0.68 & 24.66 & \textbf{0.74} \\
Italian    & \textbf{1.61}  & 0.65 & 8.13  & \textbf{0.80} \\
Japanese   & \textbf{4.98}  & 0.76 & 5.18  & \textbf{0.82} \\
Korean     & 3.74           & 0.73 & \textbf{3.23}  & \textbf{0.81} \\
Polish     & \textbf{2.36}  & 0.79 & 11.47 & \textbf{0.83} \\
Portuguese & \textbf{1.49}  & 0.73 & 5.31  & \textbf{0.83} \\
Romanian   & \textbf{8.77}  & 0.72 & 19.26 & \textbf{0.80} \\
Russian    & \textbf{2.73}  & 0.71 & 2.98  & \textbf{0.79} \\
Spanish    & \textbf{1.35}  & 0.69 & 1.60  & \textbf{0.78} \\
Thai       & \textbf{3.45}  & 0.68 & 11.10 & \textbf{0.81} \\
Turkish    & \textbf{2.75}  & 0.68 & 13.00 & \textbf{0.81} \\
Ukrainian  & 18.77          & 0.65 & \textbf{14.68} & \textbf{0.76} \\
Vietnamese & 5.12           & 0.71 & \textbf{3.88}  & \textbf{0.76} \\
\midrule
\tabletotalrow Macro-average & \textbf{7.67} & 0.71 & 10.34 & \textbf{0.79} \\
\bottomrule
\end{tabular*}

\end{table}

\section{Release and Reproducibility}
\label{app:release-details}

\paragraph{Release status.}
We have released \dataset-testset with its manifests and annotations, together
with the complete preprocessing and evaluation pipeline, enabling inspection of the benchmark construction in the demo page. We also maintain a
frozen snapshot of the 229K-hour dataset used in this paper, including its manifests, annotations,
preprocessing metadata, and processing configurations.

\paragraph{Full-dataset release.}
Upon acceptance, we will release the full 229K-hour \dataset representation used in this work, including versioned source-linked manifests, distributable annotations, metadata, and processing configurations. The released snapshot will be tied to stable manifest checksums and recorded source-access dates to support reproduction of the reported experiments.

\end{document}